\documentclass[3p,times,11pt]{elsarticle}
\makeatletter
\def\ps@pprintTitle{%
\let\@oddhead\@empty
\let\@evenhead\@empty
\def\@oddfoot{\centerline{\thepage}}%
\let\@evenfoot\@oddfoot}
\makeatother
\usepackage{float}
\usepackage{placeins}
\usepackage{graphicx}
\usepackage{amsmath}
\usepackage{amssymb}
\usepackage{enumitem}
\usepackage{color}
\usepackage{xcolor}
\usepackage{comment}
\usepackage{soul}
\usepackage{hyperref}
\usepackage{csvsimple}
\usepackage{multirow}

\newcommand{\mr}{\mathrm}
\newcommand{\mb}{\mathbf}
\newcommand{\mc}{\mathcal}

\usepackage{nomencl}
  \makenomenclature
\journal{Journal of Computational Physics, March 2022}

\begin{document}
\topmargin -1.5cm
\textheight 23cm
\begin{frontmatter}

\title{Discretization-independent surrogate modeling over complex geometries using hypernetworks and implicit representations}

\author{James Duvall}
\ead{jamesduv@umich.edu}
\author{Karthik Duraisamy}
\ead{kdur@umich.edu}
\author{Shaowu Pan}
\ead{shawnpan@umich.edu}
\address{University of Michigan, Ann Arbor, MI}


\begin{abstract}
 Numerical solutions of partial differential equations (PDEs) require expensive simulations, limiting their application in design optimization, model-based control, and large-scale inverse problems. 
Surrogate modeling techniques have the potential to decrease the computational expense while retaining dominant solution features and behavior.
Traditional Convolutional Neural Network-based frameworks for surrogate modeling require lossy pixelization and data-preprocessing, and generally are not effective in realistic engineering applications.
We propose alternative deep-learning based surrogate models for discretization-independent, continuous representations of PDE solutions, which can be used for learning and prediction over domains with complex, variable geometry and mesh topology.
Three methods are proposed and compared; design-variable-coded multi-layer perceptron (DV-MLP), design-variable hypernetworks (DV-Hnet), and non-linear independent dual system (NIDS).
Each method utilizes a main network which consumes pointwise spatial information to provide a \emph{continuous representation}, allowing predictions at any spatial location in the domain.
Input features include a minimum-distance function evaluation to \emph{implicitly encode} the problem geometry.
The geometric design variables, which define and distinguish problem instances, are used differently by each method, appearing as additional main-network input features (DV-MLP), or as hypernetwork inputs (DV-Hnet and NIDS). 
The methods are applied to predict solutions around complex, {\em parametrically-defined geometries} on {\em non-parametrically-defined meshes} with model predictions obtained many orders of magnitude faster than the full order models.
Test cases include a vehicle-aerodynamics problem with complex geometry and limited training data, with a design-variable hypernetwork performing best with the smallest error metrics and a competitive time-to-best-model despite a much greater parameter count. 
\end{abstract}

\end{frontmatter}

\section{Introduction}
\label{sec:intro}
High-fidelity numerical simulations are ubiquitous in engineering design and analysis, but are often found to be too computationally expensive in many-query applications.
Data-driven and machine learning surrogate modeling techniques offer an interesting alternative in these situations, where model accuracy may be acceptably traded for computational savings.
However, many successful methods cannot process unstructured and varying mesh topologies across the parameter space.
This limits their application to problems defined with a unique discretization, or will require interpolation in pre-processing.
This is restrictive in  problems that involve multi-scale phenomena, typical in fluid and structural mechanics, as the computational mesh will have regions of tightly-clustered points and intricate geometric features with the required interpolation to a coarser grid incurring an unacceptable loss of information and fidelity. 

Engineering models are typically expressed as systems of PDEs describing the design under consideration, with the highest fidelity versions termed the full-order model (FOM). 
Quantities of interest (QOIs) relevant to assessing the design are extracted from numerically generated FOM solutions. 
In the context of design, the model may be parameterized so that important and defining design elements may be varied, and the QOI response measured.
Often, these variations include the shape or geometry of some problem element, and these geometric design variables may be represented as  $\pmb{\mu}_{geo} \in \mathbb{R}^{n_{\mu,geo}}$.
These design variables are often used with and defined in the context of a separate parametric geometry model, relating the change in design variables to the details of the changes in the physical design. 
This geometry model is used along with a meshing tool in the process of generating FOM solutions.
The number of degrees of freedom in a FOM may be very large, and generally solutions to the FOM are expensive to obtain.
Surrogate and reduced-order modeling techniques seek to combat this computational expense.

We consider FOMs for steady-state, parametric PDEs, written generally as 
\begin{equation}
	\label{eq:generic_pde2}
	\mc{R}(\mb{x}, \mb{q}(\mb{x}), \ldots; \mb{p}_{\mc{R}} )= 0, \quad \mb{x}\in\Omega,
\end{equation}
with boundary conditions prescribed as 
\begin{equation}
	\label{eq:generic_bc2}
	\mc{B}(\mb{x},\mb{q}(\mb{x}), \ldots; \mb{p}_{\mc{B}})=0, \quad \mb{x} \in \partial \Omega, 
\end{equation}
where $\mb{q}$ is the state, $\Omega/\partial\Omega \subset \mathbb{R}^{n_{x}}$ are the problem domain/boundary, $\mb{x} \in \Omega$ are the spatial coordinates, $\mc{R}$ is the PDE operator, and $\mc{B}$ is the boundary-condition operator. 
The vectors $\mb{p}_{\mc{R}}$ and $\mb{p}_{\mc{B}}$ parameterize the governing equations and boundary conditions respectively; either \emph{explicitly} by appearing in the resulting relations or \emph{implicitly} as generative factors or geometric design variables. 
Consider $\pmb{\mu}$ to be the collection of all such parameters, written as
\begin{eqnarray}
	\label{eq:mu_def}
	\pmb{\mu} \triangleq \begin{bmatrix} \mb{p}_{\mc{R}}^T  & \mb{p}_{\mc{B}}^{T}\end{bmatrix}^{T} \in \mathbb{R}^{n_{\mu}}.
\end{eqnarray}
The geometric design variables $\pmb{\mu}_{geo}$ appear as elements of $\mb{p}_{R}$ and/or $\mb{p}_{B}$, and thus appear in $\pmb{\mu}$, which may also contain PDE coefficients or numerical values of boundary or operating conditions.
In our context, the elements of $\pmb{\mu}$ apply to the entire physical domain over which the FOM is solved, and for the problems we consider consist solely of geometric design variables.

Although we have referenced the governing PDE system, the surrogate models proposed here do not make explicit reference to the governing PDE operators: Our approach is non-intrusive and seeks to construct the predictive map using data. 
The models predict the PDE solution given only the spatial coordinates $\mb{x}$, a minimum-distance function evaluated at $\mb{x}$ to implicitly encode the geometry, and global parameters $\pmb{\mu}$.
The minimum-distance function and global parameters are each viewed as \emph{implicit representations} of the solution instance in some capacity.
Use of a minimum-distance function implies that knowledge of the geometry model is on hand, and that it may be used to generate a new design without solving the FOM.
This scenario is in contrast to DeepONet \cite{lu2021deeponet, Cai_2021, wang2021learning} and Neural Operator \cite{li2020neural,li2020multipole} methods, which are also capable of handling unstructured data without interpolation. 
However, DeepONet and Neural Operator develop mappings between input functions appearing \emph{explicitly} in the governing equations defined on the domain $\Omega$, before the operator, and the state after the operator is applied.
We use the terms \emph{parameters, design variables}, and \emph{generative factors} interchangeably, and it is important to keep in mind that our parameters $\pmb{\mu}$ are \emph{not} a function of space; they describe the entire solution instance.

The outline of this paper is as follows. 
A brief overview of relevant methods from scientific machine learning is given in Sections \ref{s:bg_podcnn} - \ref{s:bg_graph_pointcloud}, along with some recent advances regarding signed-distance function neural networks for computer graphics in Section \ref{s:sdf_networks_bg}, and hypernetworks in Section \ref{s:hypernetworks_bg}. 
Details of the proposed models are given in Section \ref{s:methods}, which include design-variable-coded multi-layer perceptron (DV-MLP), design-variable hypernetworks (DV-Hnet) and non-linear independent dual system (NIDS).
Numerical experiments and discussion are presented in Section \ref{s:results}, where the models are applied to a 2D Poisson problem and a 2D RANS problem for aerodynamic flows around complex vehicle shapes.
Concluding remarks are given in Section \ref{s:conclusions}, and supplementary information may be found in the Appendix. 

Similarities may be drawn between elements of the proposed models and existing works, but the context in which they are used and the details of the implementations are novel contributions of this work.
DV-MLP is a simple feed-forward neural network, with all input features fed through the main network, while DV-Hnet and NIDS introduce an additional hypernetwork to generate all, or only a portion, of the main-network weights and biases as a function of the design variables.
DV-MLP shares similarities with existing works, where a learned latent representation is included in addition to spatial coordinates as input features to a pointwise network \cite{Sekar2019, Park2019}.
However, DV-MLP utilizes the design variables themselves for this purpose without introducing a separate embedding network, or the complication of learning the representation concurrently during training.
NIDS shares architectural similarity with existing DeepONet models \cite{lu2021deeponet}, but the usage, requirements, and interpretation  distinguish NIDS from DeepONet, as covered in more detail in Section \ref{s:compare_nids_deeponet}.
DV-Hnet applies simple MLP hypernetworks \cite{ha2016hypernetworks} in the context of surrogate modeling, and the details of the hypernetwork and the problem scenario with complex geometry, are novel.
NIDS and DV-Hnet may be interpreted in the context of Neural Implicit Flow \cite{pan2021neural}, but the present work is a parallel effort specific to surrogate modeling with complex geometry.

Code for our model implementations may be found at \url{https://github.com/jamesduv/DISM}.
\section{Background and Related Work}
\label{s:bg_related} 
This section provides a brief overview of relevant data-driven and machine learning methods for surrogate modeling and operator regression.
Recent advances in deep learning for computer graphics are also presented, along with the concept of hypernetworks.

\subsection{Discretization-Dependent Methods}
\label{s:bg_podcnn}
As a classical dimensionality-reduction technique, proper orthogonal decomposition (POD) has been used to construct surrogate and reduced-order models \cite{dolci2016proper,salmoiraghi2018free,Willcox2002,benner2015survey}. 
Despite many attractive properties, conventional POD implementations process discretized data, and require the use of a fixed topology mesh across \emph{all parameter regimes}, fixing the number of degrees-of-freedom. 
This is restrictive in many engineering problems in which various solution features (e.g., relative motion of bodies, crack propagation, etc.) may emerge in different regions of parameter space. 
Further, data may be available from multiple sources with varying discretization and mesh topologies.
Other snapshot-based methods inherit these disadvantages, including autoencoders and their variants.

Convolutional neural networks (CNNs) have been used to construct surrogate models for both steady-state \cite{Guo2016, Bhatnagar2019, Hasnain2020,Thuerey_2020, Olaf2015Unet} and time-varying parametric problems \cite{Xu2020, Hasegawa_2020} by including an additional time-advance model such as an LSTM or temporal-convolutional network.
However, they place even greater restrictions on the discretization than POD-based methods, requiring inputs and outputs to be defined on regular Cartesian grids with consistent dimensions for all parameter regimes.
Overcoming this restriction requires interpolation from the computational mesh to a Cartesian grid overlain on the problem domain, equivalent to pixelization.
The  interpolation results in a number of undesirable effects, including a reduced-fidelity representation of the domain geometry, and a loss of information in regions of tightly-clustered mesh points, such as within boundary layers, shocks, and wakes.
The models may then be conceptualized as image-to-image mappings, where researchers Guo et al. \cite{Guo2016} note improved results when using a signed-distance field as the network input in contrast to a binarized representation.

A similar approach sidesteps the grid-regularity requirement for the output space by using a parameterization network and a separate flow-prediction network for continuous predictions of laminar airfoil flow fields, with variations in Reynolds number and angle of attack \cite{Sekar2019}. 
The parameterization network is a predictive autoencoder, used in a supervised sub-task of inferring a latent representation of the airfoil surface, requiring a consistent discretization of $x$-coordinates for all shapes. 
The flow network is a dense neural network which uses the learned latent representations and provides pointwise predictions of pressure and $x/y$ velocity components, with training errors less than 1\% and testing errors less than 3\%.
Thus the network is capable of providing continuous predictions, but still has discretization dependence due to the parameterization network. 

Another more problematic but surprisingly overlooked issue is that the memory requirements for 3D convolutions, commonly implemented on a single GPU, are not affordable for typical resolutions in realistic engineering problems. 
When taking mini-batch training into consideration, even storing the output of one single hidden layer (a 5-dimensional tensor), requires memory typically on the order of $O(10)-O(10^2)$ GB for a 3D Cartesian field with 40 million cells. 
As a result, most reported works using 3D CNNs for engineering problems are limited to below 1 million degrees of freedom \cite{santos2020poreflow}.

\subsection{Partially or Fully Discretization-Independent Methods}
\label{s:bg_graph_pointcloud}
Graph neural networks have been developed to extend  CNNs to problems defined on non-Euclidean domains, or with non-regular Cartesian structure.
In the context of modeling physical simulations, the computational mesh may be treated as a graph, $\mc{G}=(\mc{V}, \mc{E})$, where $\mc{V}$ is the set of vertices representing points in the computational domain, and $\mc{E}$ is the set of edges defining the connections among the nodes corresponding to mesh connectivity. 
Graph neural networks may be classified as either spectral \cite{Bruna2014, Henaff2015, Kipf2017, defferrard2016convolutional} or spatial \cite{duvenaud2015convolutional} approaches, although the two may be generalized by the message-passing graph neural network (MPGNN) \cite{Gilmer2017}. 
MPGNNs have been used for body-force predictions of aerodynamic flows \cite{ogoke2020graph}, and as part of the Neural Operator methods to iteratively perform the kernel integration \cite{li2020neural, li2020multipole, li2020fourier}.

Additionally, MPGNNs are used as a sub-component for certain learning and prediction schemes, with a focus on PDEs in either a mesh-based \cite{Xu2021cpnets,pfaff2021} or mesh-free scenario \cite{Sanchez-Gonzalez2020}. 
These methods operate in the computational domain and are used to advance a solution field from one time instance to the next, serving as a model for the high-fidelity simulator. 
The architectures consist of encoder-processor-decoder components, with MPGNNs used in the processor to compute interactions among computational nodes. 
In some instances, a particle-based representation of the simulation is used ~\cite{Sanchez-Gonzalez2020}, where message passing is used to capture non-local interactions between discrete particles in the simulation. 
While others adopt a more finite-volume-method inspired perspective in constructing the processor \cite{pfaff2021,Xu2021cpnets}, with message passing used to represent fluxes. 
Impressive results are seen with these methods, and they overcome many of the shortcomings of  CNN-based approaches. 
Conditionally parameterized networks \cite{Xu2021cpnets} also share some similarity with our proposed methods, in that neural network weights are treated as parametric functions, much like our design-variable hypernetworks generate the weights and biases for the main networks. 
However, the focus in those works is on simulating a particular problem instance forward in time, while here we focus on surrogate modeling of steady-state, parametrically-related cases.

Point cloud neural networks are useful in situations in which the data is available in the form of unstructured point clouds, such as the raw output of a LIDAR unit or other three-dimensional sensor. 
As such they are often seen in the context of autonomous vehicles or robotic vision. 
PointNet\cite{Qi2017PointNet} and PointNet++\cite{qi2017pointnet++} are architectures designed for point clouds and are used for scene recognition, classification, and segmentation tasks. 
The networks consume a point cloud corresponding to a 3D scan or mesh, and either offer an overall classification score for the scene, or a point-by-point segmentation score, where the goal is scene analysis. 
The network produces a global feature vector upon processing a point cloud, which is used in turn by a global classifier network, or a segmentation network. 
The segmentation network provides a pointwise score, and it is possible that a network with such an architecture may be used in a predictive setting instead. 
This is demonstrated with PointNet++ used to predict viscous, incompressible flows over 2D shapes lying on unstructured meshes  \cite{Kashefi_2021}.

Some techniques blur the line between \emph{solving} PDEs and regressing approximate PDE solutions from data.
Physics-informed neural networks~\cite{Raissi2019,sun2020surrogate} (PINNs) are an example of this.
PINNs may be seen as a modern extension of methods to solve ODEs/PDEs without data using neural networks, introduced in the late 90's \cite{Lagaris1998, Lagaris2000}.
The general idea is to embed the governing ODEs/PDEs in the loss function, and to compute the required derivatives directly from the neural-network prediction.
Doing so requires a neural network which consumes spatial coordinates $\mb{x}$, and computing derivative terms is facilitated by modern automatic-differentiation-based  deep learning packages.
However, PINNs allow the possibility of including a data term along with the governing-equation loss.
In this work, we do not include `physics-informed' terms in the loss functions, but the pointwise action of the neural networks is similar to what we propose, though the other details, such as network inputs and structure, are different.
Another class of relevant techniques capable of handling unstructured data includes operator-regression methods, such as those based on DeepONet \cite{lu2021deeponet, wang2021learning,Cai_2021}, Neural Operator \cite{li2020neural,li2020multipole}, Fourier basis networks\cite{li2020fourier}, and  GMLS Nets \cite{trask2019gmlsnets}. 
DeepONet and Neural Operator methods have shown impressive results, but as was mentioned previously, they seek to develop mappings between spatially-varying input functions appearing \emph{explicitly} in the governing equations and the solution.
Our proposed NIDS models have a similar structure as DeepONet, but their usage and requirements are different, as is expanded upon in Section \ref{s:compare_nids_deeponet}.

\subsection{Signed-Distance Function Neural Networks}
\label{s:sdf_networks_bg}
Given a set of points representing a closed surface or boundary $\partial \Omega$, the signed-distance function may be defined as 
\begin{equation}
	\label{eq:sdf_def}
	f_{\mr{sdf}}(\mb{x}; \partial \Omega) \triangleq \begin{cases}
		\phi(\mb{x}, \partial\Omega) \quad &\mb{x} \not\in \Omega \\
		0 &\mb{x} \in \partial \Omega\\
		-\phi(\mb{x}, \partial\Omega)  &\mb{x}\in \Omega
	\end{cases},
\end{equation}
where
\begin{equation}
	\label{eq:min_dist}
	\phi(\mb{x}, \partial \Omega) \triangleq \inf_{\mb{y} \in \partial \Omega} d(\mb{x}, \mb{y})
\end{equation}
is a minimum-distance function (MDF), and $d(\cdot, \cdot)$ is the Euclidean-distance function.
Stated simply, the SDF is the minimum distance between the field point and the boundary in consideration. It is positive for points outside the object ($\mb{x} \not\in \Omega$), negative for points inside ($\mb{x} \in \Omega$), and identically zero on the surface.

Recently, neural networks have been used to represent 3D objects for rendering tasks. 
The object's surface is implicitly represented as the zero-level-set of the directly regressed signed-distance field \cite{Park2019,  Davies2020} or decision boundary \cite{Chen_2019_CVPR, Mescheder_2019_CVPR}, and this problem scenario is generalized by the concept of implicit neural representations \cite{sitzmann2020implicit}.
In Ref. \cite{Park2019}, SDF predictions are made for an entire class of shapes.
This is achieved through use of auto-decoder networks, where a shape-code for each individual of the class is learned concurrently as the network is trained.
The auto-decoders consume the spatial coordinates $\mb{x}$ where the SDF prediction is required, along with the learned shape-code.
This is a powerful concept, as it allows for a single network to predict the SDF for many shapes, but it may come at the expense of blurred fine-object details.
This is noted in Ref. \cite{Davies2020}, where instead a neural network  architecture is fixed and overfit to each shape's SDF individually.
Then, to make predictions for a rendering, the weights for the shape under consideration are loaded and used.
This method may provide greater accuracy, but has the additional complications of training a separate network for each object, and for loading and unloading weights.
Additionally, this does not allow one to make SDF predictions for objects outside of the training set.

\subsection{Hypernetworks}
\label{s:hypernetworks_bg}
Hypernetworks  constitute a metamodeling approach where one neural network is used to generate the weights of another main network \cite{ha2016hypernetworks}, and are part of a broader class of proposed techniques where network weights are conditioned on model inputs or features \cite{NIPS2015_spatialtransformer, jia2016dynamic, bertinetto2016learning}.
Hypernetworks were originally applied to convolutional and recurrent neural networks for image- and natural-language-processing tasks, with the goal of reducing the number of trainable parameters while maintaining or improving model accuracy.
In such models, the weights of the main network are generated on a layer-by-layer basis, where the hypernetwork consumes a layer-embedding vector and outputs the weights for that layer.
The use-cases for hypernetworks have largely been the domain of computer science, but lately have been applied to scientific machine learning in some instances.
Pan et al. \cite{pan2021neural} leveraged hypernetworks to learn latent representation from turbulence on arbitrary meshes. 
To the best of our knowledge, it is the first time that non-linear dimensionality reduction is performed on 3D homogeneous isotropic turbulence with over 1 million cells.

A similar concept termed HyperPINNs applies hypernetworks to PINNs for parametric PDE solutions of 1D-viscous Burger's and the Lorenz system, with improved accuracy seen over baseline PINN models despite a smaller main network \cite{belbute-peres2021hyperpinn}.
The PDE parameters, which appear \emph{explicitly} in the governing equations, serve as the inputs to the hypernetwork, generating the main-network weights. 
This is very similar to what we propose in Section \ref{s:methods_dvhnet}, as design-variable hypernetworks, but there are a few key differences.
We do not use terms from the governing PDE in our loss, so use of the label ``PINN" does not apply. 
An important aspect of our methods is that local geometric features are encoded implicitly.
Additionally,  we use an MDF evaluation as an input feature in addition to the spatial coordinates.
These differences are of course related to the class of problems we aim to model, which include complex and variable geometry; a different scenario than that considered with HyperPINNs.
\section{Methods}
\label{s:methods}
In this section, we detail the proposed methods, first introducing design-variable-coded MLP and design-variable hypernetworks before introducing NIDS, which is interpreted as a partial design-variable hypernetwork. 
We begin by first describing the problem scenario and desired models in further detail.

\subsection{Problem Setup}
Denote the solution snapshot for a single instance $j$ of the FOM as 
\begin{eqnarray}
	\label{eq:dset_j}
	\mathcal{D}_{j} \triangleq \begin{Bmatrix} \begin{Bmatrix}\mb{q}_{i} \mid \mb{x}_{i} \end{Bmatrix}_{i=1}^{n_{j}}, \; \pmb{\mu}^{j} \end{Bmatrix},
\end{eqnarray}
where the solution output-input pairs are defined at $n_{j}$ spatial locations.
Considering a dataset 
\begin{eqnarray}
	\label{eq:dset}
	D \triangleq \begin{Bmatrix} \mathcal{D}_{1}, \; \mathcal{D}_2, \; ... \;, \mathcal{D}_{n_D}\end{Bmatrix}
\end{eqnarray}
containing $n_D$ snapshots, a distinctive feature of our approach is that each snapshot may correspond to a solution domain with different spatial extent and discretization, with varying number and location of mesh points. 
We seek models that can approximate the solution snapshots stored in $D$, without interpolation. 
In other words, given the generative factors or design variables for a problem $\pmb{\mu} \in \mc{M} \subset \mathbb{R}^{n_{\mu}} $, predict the system state $\mb{q}$ at any location $\mb{x}$. 
Denote the input space as $\mb{x} \in \mc{X} \subset \mathbb{R}^{n_{x}} $, and the output space as $\mb{q} \in \mc{Q} \subset \mathbb{R}^{n_{q}} $. 

The desired model should approximate the mapping $f:\mc{M} \times \mc{X} \rightarrow \mc{Q}$, without direct knowledge of the system state. 
For the problems we consider, with complex and variable geometry, the input space is augmented to include an additional minimum-distance-function coordinate, defined as
\begin{equation}
	\label{eq:xprime_def}
	\mb{x}' \triangleq \begin{bmatrix} \mb{x}^{T}, & \phi(\mb{x};\pmb{\mu})\end{bmatrix}^{T}.
\end{equation}
This defines an augmented input space, $\mc{X}'\subset \mathbb{R}^{n_{x} + 1}$, where $\mb{x}' \in \mc{X}'$, and in turn the 
desired mapping is 
\begin{equation}
	\label{eq:augmented_mapping}
	f:\mc{M} \times \mc{X}' \rightarrow \mc{Q}.
\end{equation}
The model approximation is then written as $f(\mb{x}', \pmb{\mu}) \approx \hat{\mb{q}}(\mb{x}', \pmb{\mu})$ or just $\hat{\mb{q}}$ in compact notation.
The proposed models which follow approximate the mapping of Equation \ref{eq:augmented_mapping} in different ways.

\subsection{Method 1: Design-variable-coded MLP (DV-MLP)}
This type of model is similar to those in Ref. \cite{Sekar2019} and \cite{Park2019}, except the design variables $\pmb{\mu}$ are used as additional inputs instead of a learned latent representation or shape-code.
The models are simple, and consist of a pointwise neural network which takes as inputs spatially-varying vectors $\mb{x}'$ along with the non-spatially-varying design variables $\pmb{\mu}$.
In the numerical experiments, a comparison is made against similar feed-forward neural networks where the design variables $\pmb{\mu}$ are \emph{not} included, with only $\mb{x}'$ as input; referred to simply as  MLP.
Additionally, simple network architectures are used, where the hidden state of each layer has the same dimension or number of nodes.
The neural network (denoted as $N_m$) and resulting prediction are written as
\begin{equation}
\hat{\mb{q}}(\mb{x}', \pmb{\mu})=	N_{m}(\mb{x}', \pmb{\mu}; \theta_{m}) ,
\end{equation}
where $\theta_{m}$ represents the set of weights and biases of the main network.

\subsection{Method 2: Design-variable Hypernetworks (DV-Hnet)}
\label{s:methods_dvhnet}
In design-variable hypernetworks (DV-Hnet), the weights and biases of the main network $\theta_{m}$ are generated by a hypernetwork which consumes the design variables $\pmb{\mu}$.
This concept may be considered an extension of ideas from ref. \cite{Davies2020}, where a neural network is overfit to every case in the training set separately, and the corresponding set of trained weights are loaded into the model to make predictions at inference time. 
Rather than loading weights and biases, a hypernetwork is used to generate the weights and biases for each case, and the model is trained on all cases concurrently.

The hypernetwork is written as
\begin{equation}
	\label{eq:dv-hnet_def}
	\theta_{m}(\pmb{\mu}) = N_{h}(\pmb{\mu}; \theta_{h}),
\end{equation}
and the main-network prediction written as
\begin{equation}
	 \hat{\mb{q}}(\mb{x}'; \theta_{m}(\pmb{\mu})) = N_{m}(\mb{x}';\theta_{m}(\pmb{\mu})),
\end{equation}
where $\theta_{h}$ and $\theta_{m}$ are the weights and biases of the hypernetwork and main network.
In our experiments, simple feed-forward neural networks are used for both the main network and the hypernetwork. 
All of the weights and biases contained in $\theta_{m}$ are generated at once as one large vector which is sliced and reshaped as required.
In doing so, the number of trainable parameters is increased as compared to DV-MLP, not reduced, as is the goal with many hypernetwork models.
Differing hypernetwork architectures which would result in a reduction of the number of trainable parameters are possible, but are outside the scope of this paper.
 
Note that the main network is a function of $\mb{x}'$ only. 
Experiments were performed where $\pmb{\mu}$ was also included in the main-network inputs, and the performance found to be worse by all error metrics, though these experiments were not exhaustive.
DV-Hnets are trained end to end, without over-fitting networks to each case separately, thus only the parameters in $\theta_{h}$ are adjusted during training.
To generate the predictions for a single case, the hypernetwork is first evaluated to obtain $\theta_{m}$ and the main network is composed.
Only a single forward-pass of the hypernetwork may be coupled with as many forward-passes of the main network as there are mesh points to generate a full prediction.

\subsection{Method 3: Non-linear Independent Dual System (NIDS)}
\label{s:NIDS}
\begin{figure}[H]
	\centering
	\includegraphics[width=0.75\textwidth]{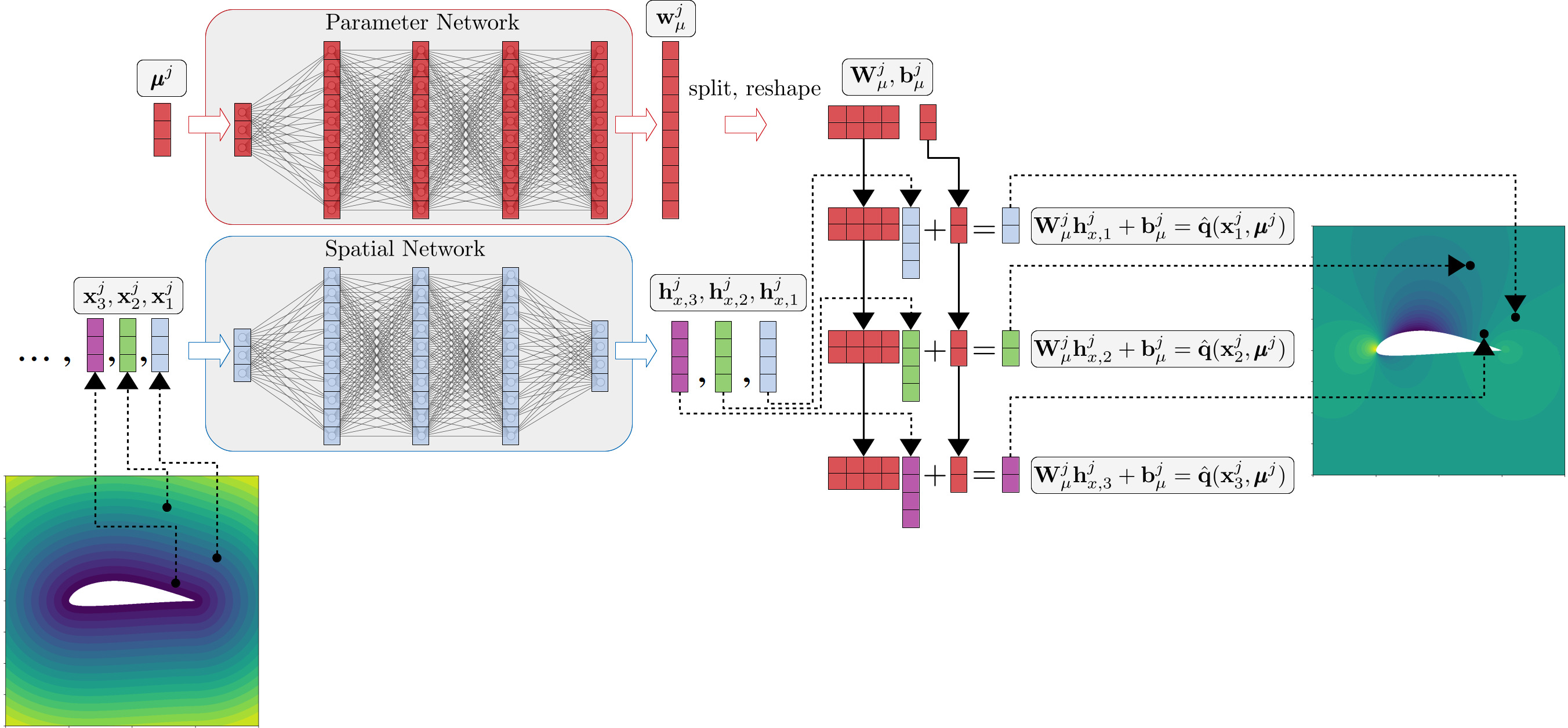}
	\caption{Schematic diagram of a NIDS network emphasizing reuse of parameter network outputs. Ref. \cite{LeNail2019} was used in making this figure.}
	\label{f:NIDS_schematic}
\end{figure}

NIDS may be conceptualized as a design-variable hypernetwork which generates only the weights and biases of the final output layer of the main network.
In this context, the hypernetwork is referred to as the parameter network.
The spatial network, which is the main network \emph{except} the output layer, and its output vector are defined as
\begin{eqnarray}
	\label{eq:spatial}
	N_{x}(\mb{x}'; \theta_{x}) \triangleq \mb{h}_{x} \in \mc{H} \subset \mathbb{R}^{n_{h}}.
\end{eqnarray}
The parameter network and its output are defined as
\begin{eqnarray}
	\label{eq:param} 
	N_{\mu}(\pmb{\mu}; \theta_{\mu}) \triangleq \mb{w}_{\mu} \in \mc{W} \in \mathbb{R}^{(n_{q} \times n_{h})+n_{q}} = \mb{W}_{\mu}, \mb{b}_{\mu}.
\end{eqnarray}
Given this, the overall prediction is written as
\begin{eqnarray}
	\label{eq:output}
	\hat{\mb{q}}(\mb{x}', \pmb{\mu}; \theta_{x}, \theta_{\mu}) \triangleq \mb{W}_{\mu}\mb{h}_{x} + \mb{b}_{\mu}.
\end{eqnarray}
In equation \ref{eq:param}, the flattened output $\mb{w}_{\mu}$ is split and reshaped appropriately to form $\mb{W}_{\mu}$ and $\mb{b}_{\mu}$.
Thus, the parameter network generates the weights and biases for the linear output layer of the main network, and the spatial network provides the final hidden state.
Alternatively,  $\mb{W}_{\mu}$ may be viewed as a basis matrix dependent on the problem parameters, while final hidden state $\mb{h}_{x}$ are the coordinates to that basis for a specific location in space $\mb{x}$. 
 $\mb{W}_{\mu}$ is reused for every spatial location where a prediction is desired, while a new $\mb{h}_{x}$ is required. 
Thus,  a prediction for a given case requires one forward-pass of the parameter network and as many forward-passes of the spatial network as there are spatial locations, much like DV-Hnet. 
Figure \ref{f:NIDS_schematic} shows a schematic diagram which emphasizes reuse of $\mb{w}_{\mu}$.
\FloatBarrier 

Further details regarding NIDS may be found in the Appendix, Section \ref{s:appendix_nids}, including
an alternate modal interpretation of NIDS predictions.

\subsubsection{Comparison of Proposed Methods}
\begin{figure}[H]
	\centering
	\includegraphics[width=1.0\textwidth]{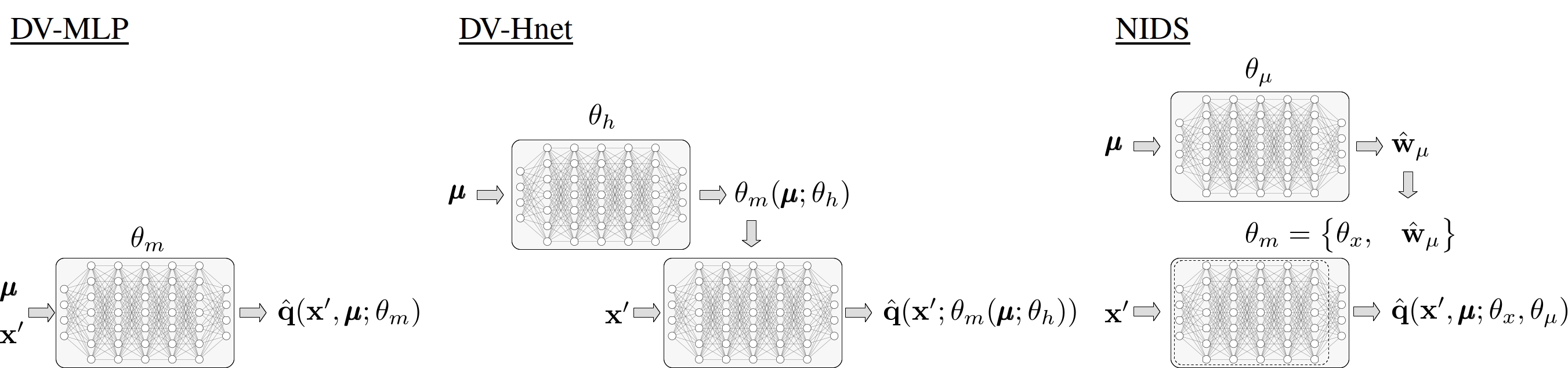}
	\caption{Schematic diagram comparing the different proposed methods. Ref. \cite{LeNail2019} was used in making this graphic.}
	\label{f:schematic_compare}
\end{figure}
\FloatBarrier 
A diagram comparing the various methods is shown in Figure \ref{f:schematic_compare}, introducing DV-MLP, DV-Hnet and NIDS models from left to right.
DV-MLP is a simple dense network, with weights and biases collected in $\theta_{m}$, set above the main-network graphic to indicate the dependence.
DV-Hnet introduces the design-variable hypernetwork and its trainable parameters $\theta_{h}$ in addition to the main network, with correspondence to the DV-MLP main network.
The hypernetwork generates the weights for the main network as a function of the design variables $\theta_{m}(\pmb{\mu})$.
In the NIDS approach,
the parameter network generates the vector $\hat{\mb{w}}_{\mu}$, and the figure highlights how the main network weights and biases $\theta_{m}$ may be considered as the spatial network weights and biases $\theta_{x}$ in addition to vector $\hat{\mb{w}}_{\mu}$.
Additionally, the dashed line in the main network corresponds to the NIDS spatial network, highlighting the slight difference between the two in the context of the other models.
Details on the implementation and training methods are provided in the Appendix, Section \ref{s:training_methods}.

\subsubsection{Comparing NIDS and DeepONet}
\label{s:compare_nids_deeponet}
DeepONet is a powerful method for operator approximation that has been developed recently \cite{lu2021deeponet}.
NIDS and unstacked DeepONet share architectural similarities, but the requirements and usage for each are different.
Essentially, DeepONet approximates the PDE operator, whereas NIDS approximates the PDE solution.
Both models consist of a pair of neural networks whose outputs are combined in a final linear layer.
The DeepONet trunk network corresponds to the NIDS spatial network,  both consuming spatial coordinates, and the DeepONet branch network corresponds to the NIDS parameter network. 
Structurally, when the predicted state is one-dimensional, the two models are identical. 
The original DeepONet does not have the capability to predict multi-dimensional states, and the output transformation is not viewed in the context of combining a weight matrix with a final hidden state.
Rather, the form of the DeepONet model was chosen such that it may be viewed analogously to a theorem regarding approximation of operators by neural networks \cite{chen1995universal}.
Additionally, both the trunk and branch networks consume spatially varying quantities, with the branch network consuming input functions or parameters, while only the spatial network of NIDS consumes such inputs.
This is a fundamental difference regarding the spaces between which each network acts.
Further, the branch network requires a sampling of input functions at consistent points in space spanning the domain over which predictions are to be made.
No such usage or requirement exists for the NIDS parameter or spatial networks, and such a requirement would break the discretization independence sought here.

\section{Numerical Experiments}
The proposed models of Section \ref{s:methods} are compared on a 2D Poisson problem and a 2D RANS problem for vehicle aerodynamics in this section.
First, error metrics and model architectures are defined in Sections \ref{s:error_metrics} and \ref{s:model_architectures}, where the main-network architecture is kept fixed among the various models.
Results and discussions are provided in Sections \ref{s:results_poisson} and \ref{s:results_gm} for the Poisson and RANS problems respectively, and an overview of the effect of mixed-precision training is given in Section \ref{s:profiling}.
\label{s:results}
\subsection{Error Metrics}
\label{s:error_metrics}
The following error metrics are reported for the experiments:
The root-mean-squared-error (RMSE) for the $k$th component of the state is computed as 
\begin{equation}
	\label{eq:2dshapes_mse}
	\mr{RMSE}_{k} \triangleq \sqrt{\frac{1}{N}
	\sum_{j=1}^{n_{c}} \sum_{m=1}^{n_{j}} \bigg( \hat{\mb{q}}_{k}(\mb{x}_{m}^{j}, \pmb{\mu}^{j};\theta) - \mb{q}_{k}(\mb{x}_{m}^{j}, \pmb{\mu}^{j}) \bigg)^{2}},
\end{equation}
where
\begin{equation}
	N = \sum_{j=1}^{n_{c}} n_{j}
\end{equation}
is the total number of points in all $n_{c}$ cases, where there are $n_{j}$ points in the $j$th case.
The mean-absolute-error (MAE) is computed analogously as
\begin{equation}
	\label{eq:2dshapes_mae}
	\mr{MAE}_{k} \triangleq \frac{1}{N} \sum_{j=1}^{n_{c}} \sum_{m=1}^{n_{j}} | \hat{\mb{q}}_{k}(\mb{x}_{m}^{j}, \pmb{\mu}^{j};\theta) - \mb{q}_{k}(\mb{x}_{m}^{j}, \pmb{\mu}^{j}) |.
\end{equation}
The RMSE and MAE are computed pointwise across all cases, meaning that every point in every case is weighted equally, regardless of the number of grid points which varies from mesh to mesh.
Both have units consistent with the predicted quantities, making them more intuitive than MSE alone.
They provide similar measures, though the RMSE penalizes larger errors more than the MAE.
One may also consider a pointwise relative error or mean-absolute-percent-error, but these metrics are problematic when the predicted field variables are very small and/or go through 0.
Instead of this, a casewise mean-relative-L2-error (ML2E) is also reported. 
To define this, gather all predictions for case $j$ in matrix $\hat{\mb{Q}}^{j}$, and all ground truth in matrix $\mb{Q}^{j}$, where $\hat{\mb{Q}}^{j},\mb{Q}^{j} \in \mathbb{R}^{n_{j} \times n_{q}}$.
Then $\hat{\mb{Q}}^{j}[:,k]$ is the full snapshot for the $k$th component of the predicted state, for the $j$th case.
Then the ML2E is computed as
\begin{equation}
    \mr{ML2E}_{k} \triangleq \frac{1}{n_{c}}\sum_{j=1}^{n_{c}} \frac{\| \hat{\mb{Q}}^{j}[:,k] - \mb{Q}^{j}[:,k]    \|_{2}}{\| \mb{Q}^{j}[:,k]  \|_{2}}.
\end{equation}
The ML2E may be multiplied by 100 and interpreted loosely as a mean percentage error.

\subsection{Model Architectures}
\label{s:model_architectures}
\begin{table}[H]
	\begin{center}
		\caption{Details on the structure of the networks used in the numerical experiments.}
		\label{t:model_architectures}
		\begin{tabular}{|c|c|c|c|c|c|}
			\hline
			 & \multicolumn{3}{c|}{ Main/Spatial Network} & \multicolumn{2}{c|}{Hyper/Parameter Network}\\
			 \hline
			Method 			& \# Hidden Layers & \# Nodes/Layer & $\mr{dim}(\mb{h}_{x})$ & \# Hidden Layers & \# Nodes/Layer \\
			\hline
			MLP/DV-MLP 		& 5 & 50 & - & - & -   \\
			DV-Hnet 		& 5 & 50 & - & 5 & 50  \\
			NIDS 			& 4 & 50 & 50 & 5 & 50 \\
			\hline
		\end{tabular}
	\end{center}	
\end{table}

\begin{table}[H]
	\begin{center}
		\caption{Summary of the network inputs and outputs for the various model types considered, along with the number of trainable parameters for the two problems considered in numerical experiments. }
		\label{t:model_io}
		\begin{tabular}{|c|c|c|c|c|c|c|}
			\hline
			& \multicolumn{2}{c|}{ Main/Spatial Network} & \multicolumn{2}{c|}{Hyper/Parameter Network} & \multicolumn{2}{c|}{\# Weights}\\
			\hline
			Method 			& inputs 	& outputs 			& inputs & outputs  & Poisson & RANS \\
			\hline
			MLP				& $\mb{x}'$ & $\hat{\mb{q}}$ & - & -                        & 10,451    & 10,553 \\
			DV-MLP 			& $\mb{x}', \; \pmb{\mu}$ &  $\hat{\mb{q}}$ & - & -         & 10,651    & 10,953\\
			DV-Hnet			& $\mb{x}'$ & $\hat{\mb{q}}$ & $\pmb{\mu}$ & $\theta_{m}$   & 543,451   & 548,853  \\
			NIDS 			& $\mb{x}'$ & $\mb{h}_{x}$ & $\pmb{\mu}$ & $\mb{w}_{\mu}$   & 23,451    & 28,853\\
			\hline
		\end{tabular}
	\end{center}	
\end{table}
A summary of the model architectures used in the following experiments is given in Table \ref{t:model_architectures}, with details on model inputs, outputs, and number of trainable parameters given in Table \ref{t:model_io}.
The main network architecture is fixed and consistent across the different model types. 
Note that this is true even for NIDS, where the final linear output layer of the main network is built using the outputs of the spatial and parameter networks.
Differences in the number of predicted field variables $n_{q}$ and the number of parameters $n_{\mu}$ causes the slight difference in the network sizes between problems.
The number of trainable parameters in DV-Hnet models is much greater than the other models, with most of the weights in the output layer of the hypernetwork. 
This is expected, as the hypernetwork output-layer weight matrix has dimension $\mr{dim}(\theta_{m}) \times n_{\mr{nodes}}$, which is $10,451 \times 50=522,550$ for the Poisson problem and $10,553 \times 50 = 527,650$ for the RANS problem, where $\mr{dim}(\theta_{m})$ corresponds to the MLP entry in Table \ref{t:model_io}.
This is an effect of generating all main-network weights at once, and other hypernetwork architectures which result in fewer trainable parameters are possible.
Such hypernetworks could generate the weights on a layer-by-layer basis or otherwise exploit structures present in the generated weights and biases.
\subsection{2D Poisson Equation}
\label{s:results_poisson}
\subsubsection{Problem Description}
\begin{figure}[h!]
	\centering
	\includegraphics[width=1.0\textwidth]{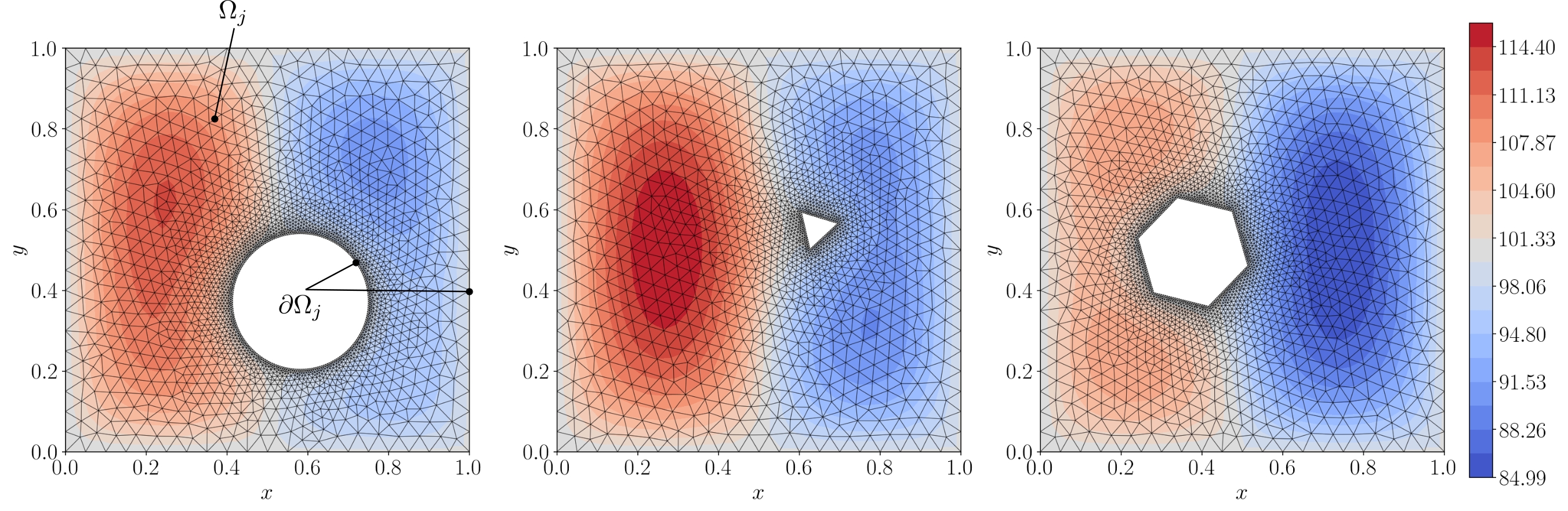}
	\caption{Example solution fields and meshes for three randomly generated shapes.}
	\label{f:2dshapes_ex}
\end{figure}

The Poisson equation with a source term is solved on a unit square two-dimensional domain with a randomly sized and oriented shape within the domain acting as another internal boundary.
While the geometry of the domain is parametric, the meshes are not, with each mesh having different spatial extent, number of points, and topology.
The square domain and its boundary (without the internal shape) are written as
\begin{equation}
	\label{eq:2dshapes_dom1}
	B = \begin{Bmatrix}x,y \mid x,y \in [0,1]  \end{Bmatrix},
\end{equation}
and
\begin{equation}
	\label{eq:2dshapes_b1}
	\partial B = \begin{Bmatrix} x,y \mid x=0, \; \mr{or} \; x=1, \; \mr{or} \; y=0, \; \mr{or} \; y=1   \end{Bmatrix},
\end{equation}
respectively.
Since each shape is different, the problem domains are also different. 
Let $\partial S_{j}$ be the set of points defining the $j$th internal-shape boundary. 
Then let $S_j$ represent the set of points which are either on the $j$th shape surface or enclosed by it. 
With this, the domain for the $j$th problem is 
\begin{eqnarray}
	\label{eq:2dshapes_dom2}
	\Omega_{j} &=& B \cap \big(S_{j} - \partial S_{j}    \big)^{C},
\end{eqnarray}
with overall boundary given by
\begin{eqnarray}
	\label{eq:2dshapes_boundary}
	\partial \Omega_{j} &=& \partial B \cup \partial S_{j}.
\end{eqnarray}
The governing Poisson equation and source term are  defined as
\begin{eqnarray}
	\label{eq:2dshapes_gov}
	\nabla^2  q(x,y) &=& f(x,y), \quad x,y \in \Omega_{j}\\
	\label{eq:2dshapes_src}
	f(x,y) &=& 
	\begin{cases}
		+500 \quad x < 0.5 \\
		-500 \quad x \geq 0.5
	\end{cases}
\end{eqnarray}
with Dirichlet boundary conditions specified for all boundaries as 
\begin{eqnarray}
	\label{eq:2dshapes_dirichletBC}
	q(x,y) &=& 100, \quad x,y \in \partial \Omega_{j}.
\end{eqnarray}
Eight classes of shapes are considered, consisting of circles and polygons with 3-9 sides. 
1000 instances of randomly scaled, located, and rotated shapes of each class are defined.  
Each shape is specified by its center point, the radius of its circumcircle (which is also the distance between the center point and each vertex), and a rotation angle measured positive counterclockwise from the $x$-axis. 
Example meshes and solution fields are shown in Figure \ref{f:2dshapes_ex}. 
The ranges for each design variable are given in Table \ref{t:poisson_mu} and these entries are used to populate parameter vector, shown in Equation \ref{eq:mu_poisson}.
The largest mesh contains 2677 points, while the smallest contains just 1341.
\begin{table}[H]
	\begin{center}
		\caption{Description of design variables for the 2D Poisson problem. }
		\label{t:poisson_mu}
		\begin{tabular}{|c|c|c|c|}
			\hline
			Symbol & Description & Range & Units \\
			\hline
			$x_{\mr{cen}}$ & $x$-coordinate of shape center point & $[0.3, 0.7]$ & -\\
			$y_{\mr{cen}}$ & $y$-coordinate of shape center point & $[0.3, 0.7]$ & -\\
			$r$ & Shape radius & $[0.05, 0.2]$  & -\\
			$\gamma$ & Shape rotation angle & $[-\pi, \pi)$ & radians \\
			\hline
		\end{tabular}
	\end{center}	
\end{table} 
\begin{equation}
	\label{eq:mu_poisson}
	\pmb{\mu} = \begin{bmatrix} x_{\mr{cen}}, & y_{\mr{cen}}, & r, & \gamma  \end{bmatrix}^{T}
\end{equation}
Further details on the solution procedure are given in Section \ref{s:appendix_poisson} of the Appendix.
The parameter vector $\pmb{\mu}$ is `incomplete,' as the class of the shape (circle, triangle, etc..) is not given.
This information enters the main network indirectly through the MDF coordinate present in $\mb{x}'$.
\FloatBarrier

\subsubsection{Results and Discussion}
Models corresponding to the architectures given in Table \ref{t:model_architectures} were trained using solution field data from all shape classes.
80\% of the 8000 available solutions are used as the training set, while the remaining 20\% are withheld in the validation set, with an equal number of training cases used from each shape class.
The training and validation groups are selected randomly and are the same for all models.
Adam optimizer was used with a learning rate of $1\times10^{-4}$ and default settings otherwise.
A batch size of 1500 points is used, resulting in 8765 batches/gradient updates per epoch.

\begin{figure}[h!]
	\centering
	\includegraphics[width=1.0\textwidth]{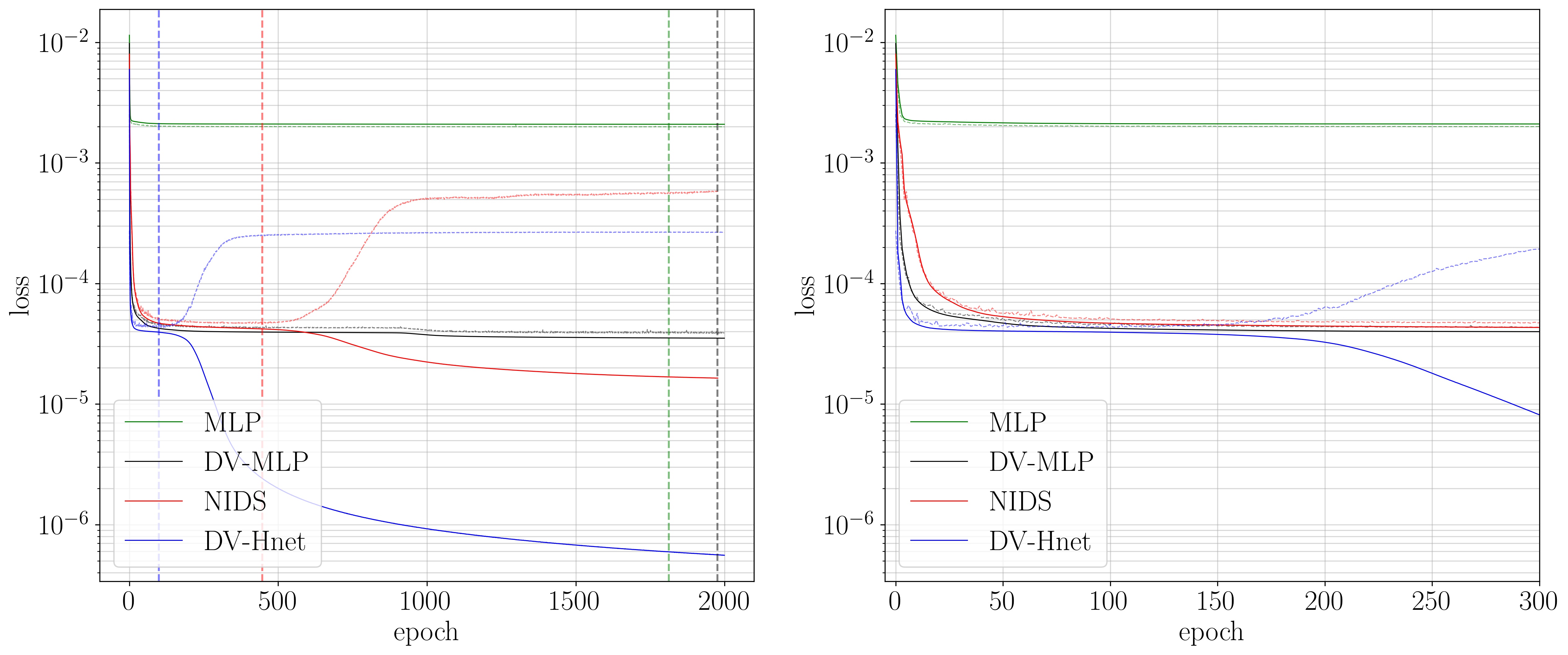}
	\caption{Comparison of training curves for the various models. The solid lines are training loss, while dashed lines are validation loss. The vertical dashes lines locate the minimum validation loss, and these weights are tracked and used in the predictions which follow.}
	\label{f:poisson_training_curves}
\end{figure}

Training curves for the various models are shown in Figure \ref{f:poisson_training_curves}, with solid lines for training loss and transparent dashed lines for validation loss. 
Vertical dashed lines correspond to the minimum validation loss, and these weights are used in generating model predictions.
Summaries of error metrics for each model type are given in Table \ref{t:poisson_errors}.
The training curves correspond to normalized quantities, with all network inputs and outputs min-max normalized to lie in the range $[0,1]$, while the errors reported in \ref{t:poisson_errors} correspond to dimensional quantities, despite the fact that units are not specified for the problem.
Table \ref{t:poisson_errors} summarizes profiling characteristics for each model.
The time-to-best-model is an estimated wall-time extrapolated from the profiling results, computed as 
\begin{equation}
\label{eq:timetobestmodel}
\mr{Time \; to\; best\; model} = \mr{Average\; Step \; Time} \times \mr{\# batches} \times \mr{\# epochs \;to \;best \; model}.
\end{equation}
A comparison of mixed and double-precision training characteristics is given later in Section \ref{s:profiling}, and for this problem mixed precision for DV-MLP and NIDS resulted in increased average step times, with small memory savings.
\FloatBarrier

\begin{table}[H]
	\begin{center}
		\caption{Summary of training and validation error metrics RMSE, MAE, and ML2E for the Poisson problem.}
		\label{t:poisson_errors}
		\begin{tabular}{|c|c|c|c|}
			\hline
			Network Type & RMSE  (train / val) & MAE (train / val)          &  ML2E (train / val) \\
			\hline
			MLP         & 1.53 / 1.49           & 1.00 / 0.97               & $1.46\cdot10^{-2}$ / $1.42\cdot10^{-2}$   \\
			DV-MLP      & 0.20 / 0.21           & 0.117 / 0.123             & $1.64\cdot10^{-3}$ / $1.72\cdot10^{-3}$  \\
			NIDS 	    & 0.22 / 0.23           & 0.137 / 0.144             & $1.87\cdot10^{-3}$ / $1.96\cdot10^{-3}$   \\
			DV-Hnet 	& 0.21 / 0.22           & 0.126 / 0.132             & $1.74\cdot10^{-3}$ / $1.83\cdot10^{-3}$    \\
			\hline
		\end{tabular}
	\end{center}	
\end{table} 

\begin{table}[H]
	\begin{center}
		\caption{Comparison of maximum memory usage, average step times during training, and an estimated time-to-best-model for the Poisson problem.}
		\label{t:poisson_profile}
		\begin{tabular}{|c|c|c|c|c|}
			\hline
			Network Type    & Max. Memory & Average Step Time       & Epoch Best    & Time to Best Model \\
			\hline
			DV-MLP          & 812 MB      & 2.9 ms                  & 1976          & 50,227 s   \\
			NIDS            & 821 MB      & 5.0 ms                  & 446           & 19,546 s   \\
			DV-Hnet         & 874 MB      & 5.8 ms                  & 99            & 5,033 s    \\ 
			\hline
		\end{tabular}
	\end{center}	
\end{table} 

Observing the training curves alone, it is apparent that MLP models which do not consume the design variables $\pmb{\mu}$ are not nearly as effective as the others, and this is supported by Table \ref{t:poisson_errors}.
Beyond this, the behavior of DV-Hnet and NIDS during training is quite different than that of DV-MLP.
The curves in the early stages of training are similar, with an initial region of rapid decrease followed by a plateau, but NIDS and DV-Hnet then proceed to overfit the training data, with large gaps developing between the training and validation losses.
While this overfitting is not desirable for extrapolation to unseen cases, it indicates increased expressiveness of the hypernetwork based models as compared to DV-MLP, analogous to what was observed by Davies et. al \cite{Davies2020}, where finer part details were resolved when networks intentionally overfit to one shape.
In practice this overfitting may be used as a marker for early-stopping, with validation-loss-divergence as the indicator.

\begin{figure}[h!]
	\centering
	\includegraphics[width=0.60\textwidth]{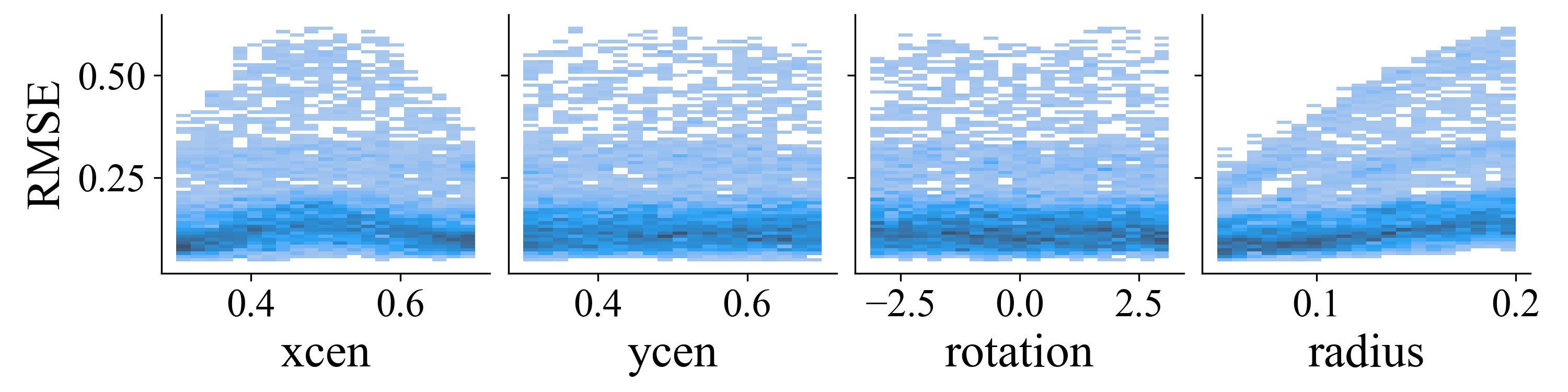}
	\caption{DV-MLP: Examining relation between RMSE and the design variables.}
	\label{f:poisson_dvmlp_hist_genfac}
\end{figure}

The divergence of the DV-Hnet validation curve occurs well before NIDS, and DV-Hnet overfits the training data to a much greater extent.
This is likely an effect of having all weights generated by a hypernetwork instead of just the output-layer weights.
All models show a small gap between the training and validation loss metrics, indicating good generalization, and all models perform similarly.
The mean percentage error is less than 1\% for all models which consume $\pmb{\mu}$.
Table \ref{t:poisson_profile} reveals that the time-to-best-model is much less for DV-Hnet than the other models due to the quicker convergence.
However, observing the training curve on the right of Figure \ref{f:poisson_training_curves} shows that around epoch 99, where the best DV-Hnet model occurs, there is only a small gap among the models.
Thus some accuracy in the DV-MLP and NIDS models could be traded for a reduced time-to-best-model if all model training was stopped at that point.
\begin{figure}%
    \centering
        {{\includegraphics[width=7cm]{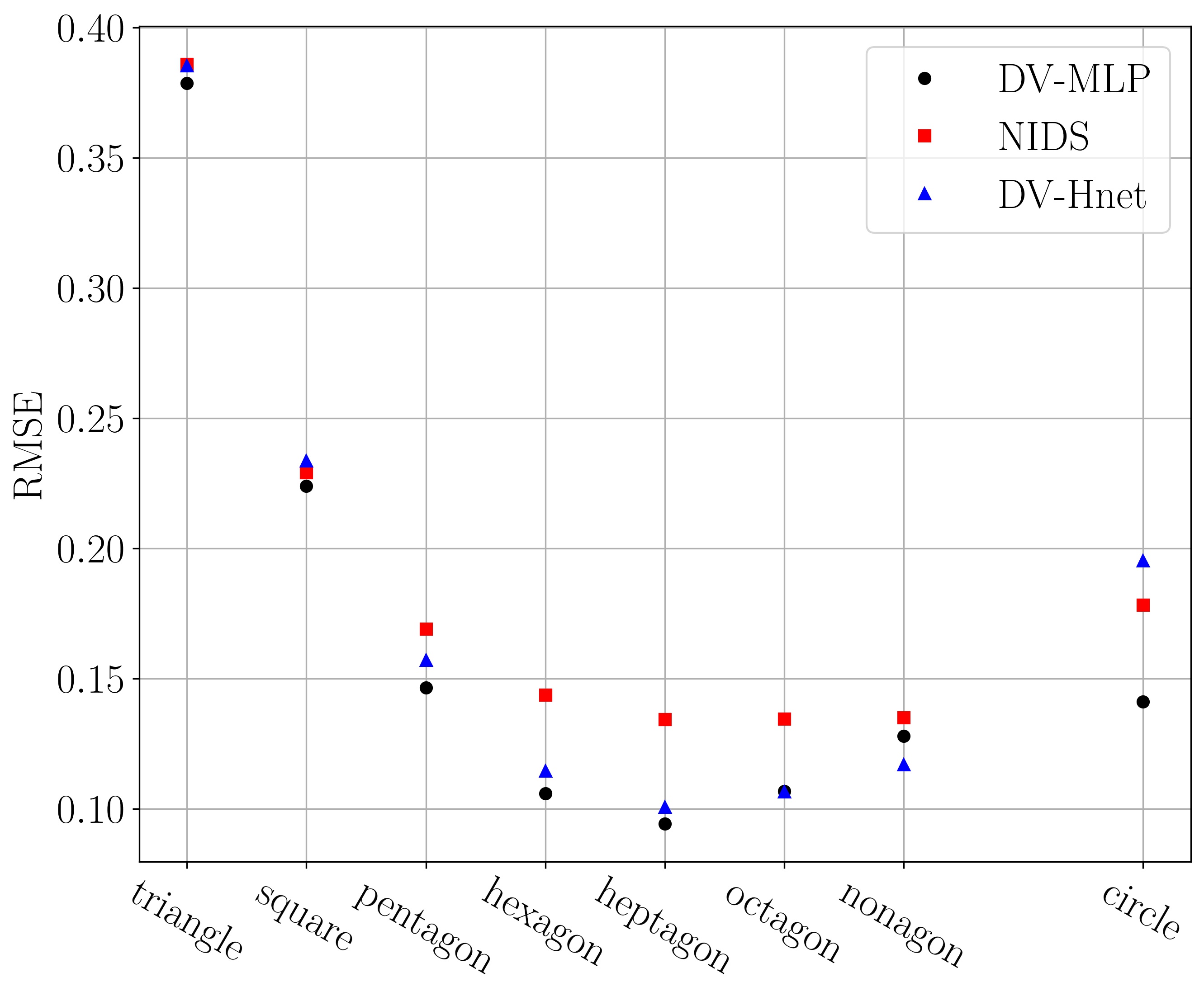} }}%
        \qquad
        {{\includegraphics[width=7cm]{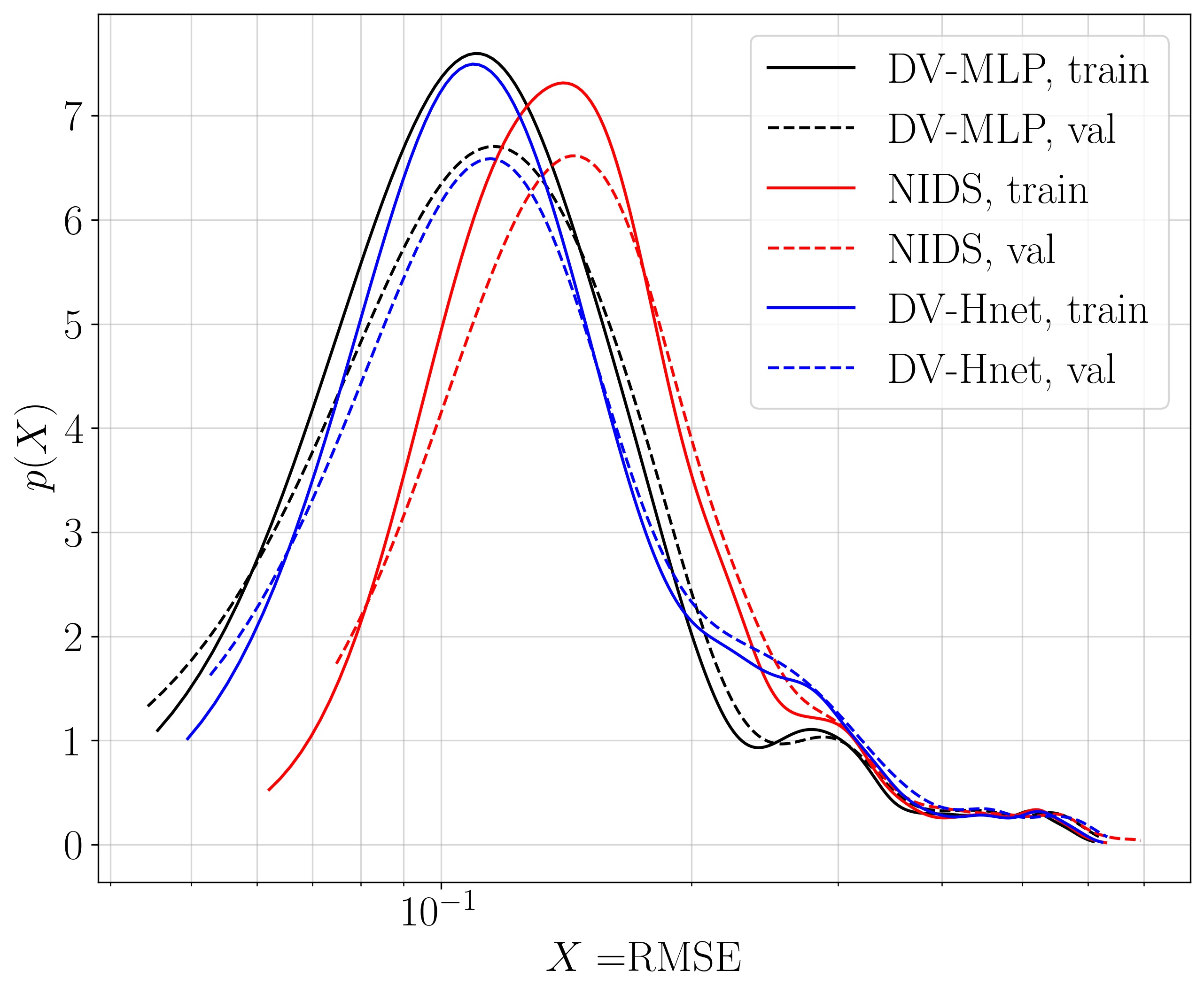} }}%
        \caption{Left: Comparing the overall RMSE (over training and validation groups) versus shape class. Right: Comparing kernel density estimates of the training and validation RMSE for each model type.}%
        \label{f:poisson_errfig}%
\end{figure}
\FloatBarrier
The variation in RMSE against the shape classes is shown on the left of Figure \ref{f:poisson_errfig}, with the general trend consistent across model types.
The model predictions are much worse for the triangles than the other shapes, and generally worse for polygons with a fewer number of sides, though this is not absolute.
Kernel density estimates for the RMSE distributions are shown on the right of Figure \ref{f:poisson_errfig}, with solid lines for the training set and dashed lines for validation.
The trends in these curves are consistent with the errors reported in Table \ref{t:poisson_errors}, with the validation distributions peaking slightly to the right of the corresponding curves for the training data.
The curves for DV-MLP and DV-Hnet are very similar to another, but with DV-Hnet having a heavier tail, leading to worse error metrics.
Further, the distribution of error metrics against the design variables may be visualized using 2D histograms, as shown in Figure \ref{f:poisson_dvmlp_hist_genfac} for DV-MLP.
The histograms for the other models are very similar and are not shown.
This implies no clear correlation between the error metrics and the elements of $\pmb{\mu}$, with the possible exception of the shape radius, where a positive trend is seen between the error metrics and increasing radii.

\begin{figure}[h!]
	\centering
	\includegraphics[width=1.0\textwidth]{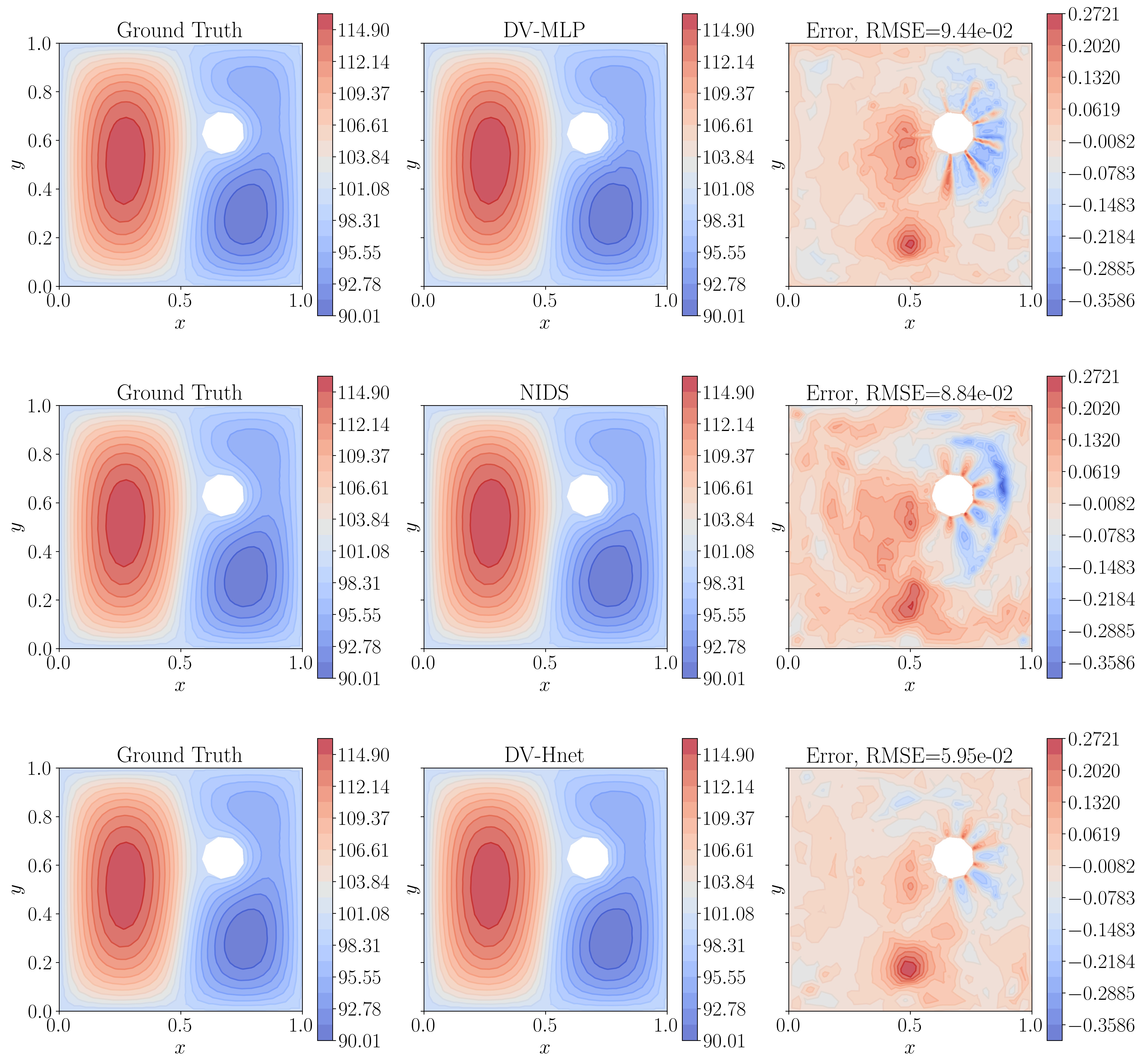}
	\caption{Comparing model predictions, unseen nonagon.}
	\label{f:poisson_nonagon_pred}
\end{figure}

Predictions on an unseen nonagon are shown in Figure \ref{f:poisson_nonagon_pred} for the various models.
Generally the predictions match the ground truth quite well and the main features of the solution field are captured.
The reported RMSE are in line with Figure \ref{f:poisson_errfig} for nonagons, with DV-Hnet performing best, and NIDS/DV-MLP performing similarly.
The structure of the error fields is similar for all models, with regions of positive errors emanating from vertices, and regions of negative error near flat sides.

\begin{figure}[h!]
	\centering
	\includegraphics[width=1.0\textwidth]{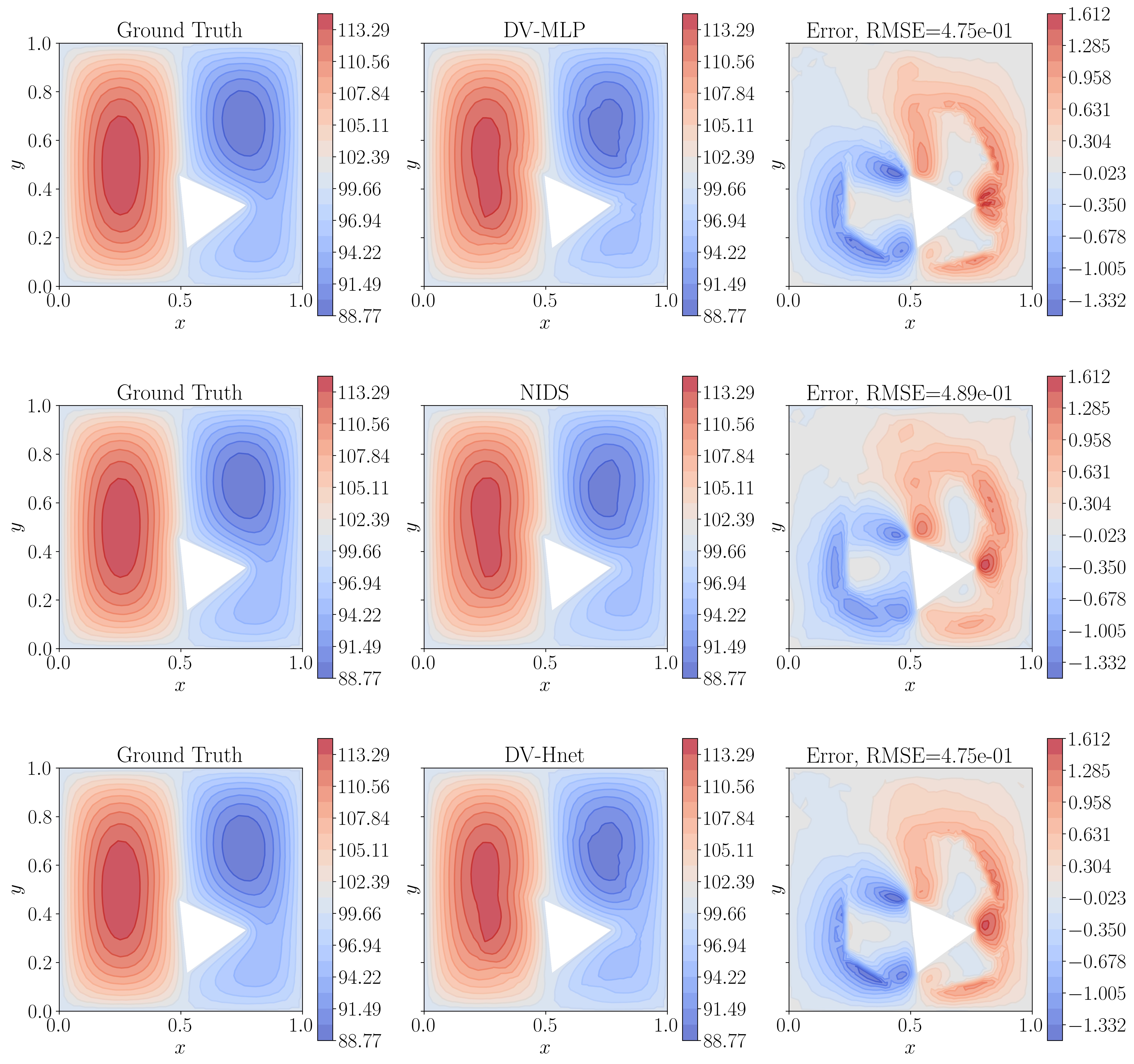}
	\caption{Comparing model predictions, unseen triangle.}
	\label{f:poisson_triangle_pred}
\end{figure}

Predictions on an unseen triangle are shown in Figure \ref{f:poisson_triangle_pred}.
As with the nonagon, the main features of the solution field are well captured by the models, but discrepancies in the contours are observed, with some twisting and distorting of contour lines in the predictions.
The reported RMSE for the triangle predictions are much greater than for the nonagon, and all models perform similarly by this metric.
The structure of the error contours is again similar among the models, with regions of larger errors emanating from the triangle's vertices.
\FloatBarrier
The similar performance of the models and the structure of the error contours are likely related to the problem formulation.
Details of the shape class enter only through the MDF coordinate $\phi(\mb{x};\pmb{\mu})$ in $\mb{x}'$, since the parameter vector $\pmb{\mu}$ does not encode shape class.
However the MDF is not a perfect feature for describing the geometry. 
In particular, larger errors are seen to emanate radially outward from the vertices of the polygons. 
In these locations, the MDF alone is insufficient as the solution at that point depends on the details of the shape surface adjacent to the vertex.
Thus it may be beneficial to include MDF spatial-gradient components as additional features in $\mb{x}'$, such as $\frac{\partial \phi}{\partial x}$ and $\frac{\partial \phi}{\partial y}$, to provide more geometric information. 
However, obtaining these features will require additional steps in pre-processing, and this trade-off must be considered.

\subsection{2D Compressible RANS, Vehicle Aerodynamics}
\label{s:results_gm}
\subsubsection{Problem Description}
\begin{figure}[h!]
	\centering
	\includegraphics[width=1.0\textwidth]{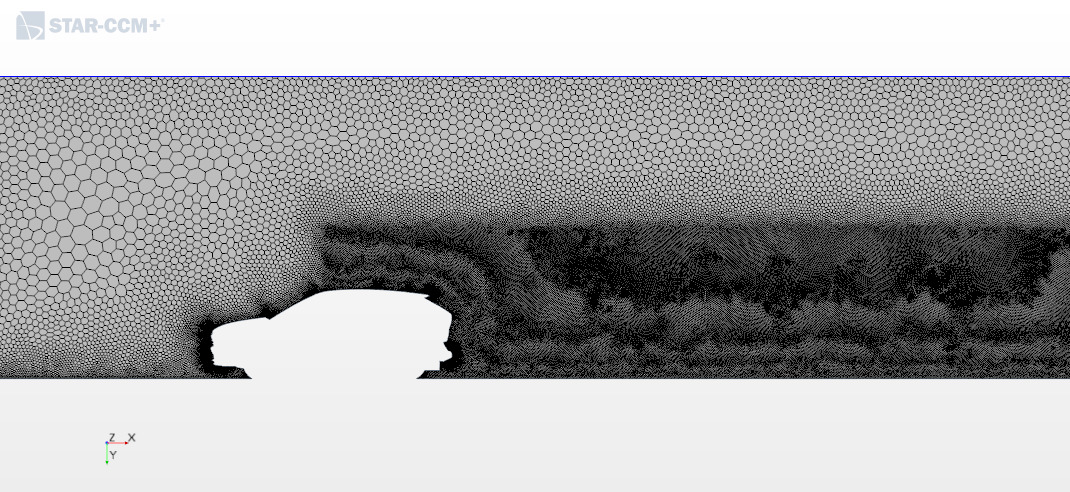}
	\caption{Example unstructured mesh surrounding a vehicle. }
	\label{f:veh_mesh}
\end{figure}

The Reynolds Averaged Navier Stokes (RANS) equations are derived by ensemble averaging the Navier Stokes equations and substituting the Reynolds-decomposed state variables. 
This decomposition separates the state variables into mean (ensemble averaged) and fluctuating components $q = \overline{q} + q^{'}$, where $q$ is a generic state variable, $\overline{q}$ is the mean, and $q^{'}$ is the fluctuating component. 
In the incompressible limit, the steady RANS equations may be written in the form
\begin{eqnarray}
	\label{eq:inc_cont}
	\nabla \cdot \overline{\mb{u}} &=& 0 \\
	\label{eq:inc_momentum}
	\rho \overline{u}_{j} \frac{\partial \overline{u}_{i}}{\partial x_{j}} &=& \frac{\partial}{\partial x_{i}} \bigg[\mu \bigg( \frac{\partial \overline{u}_{i}}{\partial x_{j}} + \frac{\partial \overline{u}_{j}}{\partial x_{i}}    \bigg) - \overline{p}\delta_{ij} - \rho \overline{u_{i}^{'}u_{j}^{'}} \bigg],	
\end{eqnarray}
where $\mb{u} = \begin{bmatrix}u & v & w \end{bmatrix}^{T}$; the $x$-, $y$-, and $z$-components of velocity. 
The incompressible RANS equations were solved using Star CCM+ with the $k$-$\epsilon$ turbulence model.
The dataset - generated by General Motors, Inc. - consists of 2D slices along the vehicle centerline for 124 unique vehicle shapes at a vehicle speed of $90$-kilometers-per-hour.
The simulations utilize unstructured meshes of varying size, with an example mesh slice shown in Figure \ref{f:veh_mesh}. 
\FloatBarrier
As with the Poisson dataset, the domain for each problem is different due to the differing vehicle shapes.
The rectangular domain, without a vehicle, may be written as 
\begin{eqnarray}
	\label{eq:gm_rect_dom}
	B &=& \begin{Bmatrix}x,y \mid x \in [-3.63, 17.7], \; y \in [0, 5.4] \end{Bmatrix}.
\end{eqnarray}  
Let $\partial S_{j}$ be the set of points which define the vehicle shape, and let $S_{j}$ be the set of points which are on the $j$th vehicle shape or enclosed by it. 
Then the problem domain and boundary are defined analogously to Equations \ref{eq:2dshapes_dom2} and \ref{eq:2dshapes_boundary}.

\begin{figure}[H]
	\centering
	\includegraphics[width=0.75\textwidth]{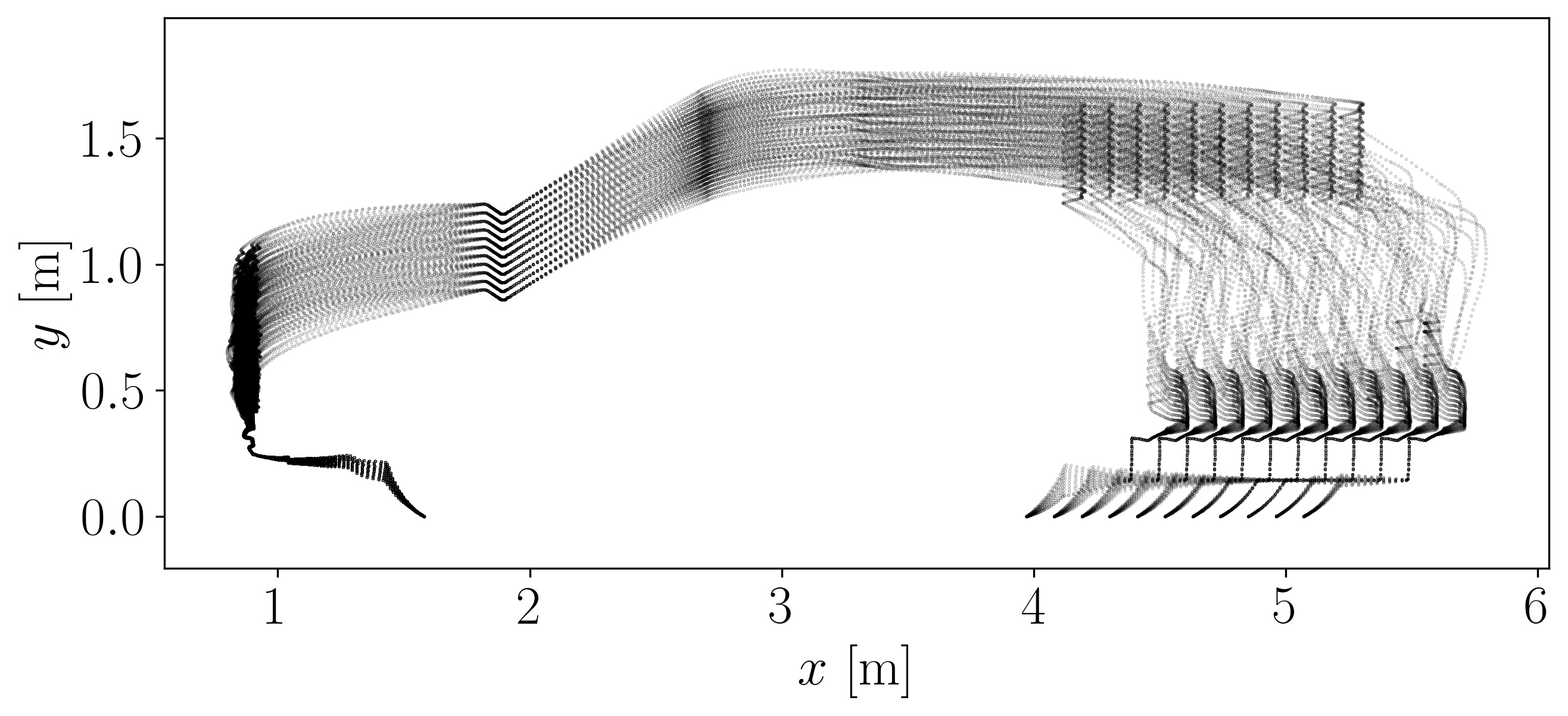}
	\caption{Composite image of 124 vehicle shapes overlain on the same axes.}
	\label{f:veh_overlay}
\end{figure}

The design variables used to generate a vehicle shape are summarized in Table \ref{t:vehicle_mu}, with all 124 shapes overlain on one set of axes in Figure \ref{f:veh_overlay}. 
This shows the wide variety of vehicle lengths and heights in the dataset, though they are all clearly related via the parameterization. 
In the following experiments, the domain was truncated slightly to include only points with $\phi(\mb{x};\pmb{\mu}) \leq 10$, resulting in mesh sizes ranging from 95,027 to 99,811.
\begin{table}[H]
	\begin{center}
		\caption{Description of entries in $\pmb{\mu}$ vector for the vehicle aerodynamics dataset.}
		\label{t:vehicle_mu}
		\begin{tabular}{|c|c|c|}
			\hline
			Design Parameter    & Units     & Range \\
			\hline
			Backlight Angle     & Degrees   & $[25, 57]$\\
			Windshield Angle    & Degrees   & $[57,63]$ \\
			Face Lip Angle      & Degrees   & $[0, 5]$ \\
			Hood Front Angle    & Degrees   & $[10, 20]$ \\
			Angle of Approach   & Degrees   & $[15, 25]$ \\
			Angle of Departure  & Degrees   & $[15, 25]$ \\
			Vehicle Length      & mm        & $[3800, 4900]$ \\
			Floor to Roof Height & mm       & $[1448, 1788]$ \\
			\hline
		\end{tabular}
	\end{center}	
\end{table} 

\FloatBarrier

\subsubsection{Results and Discussion}
\begin{figure}[h!]
	\centering
	\includegraphics[width=1.0\textwidth]{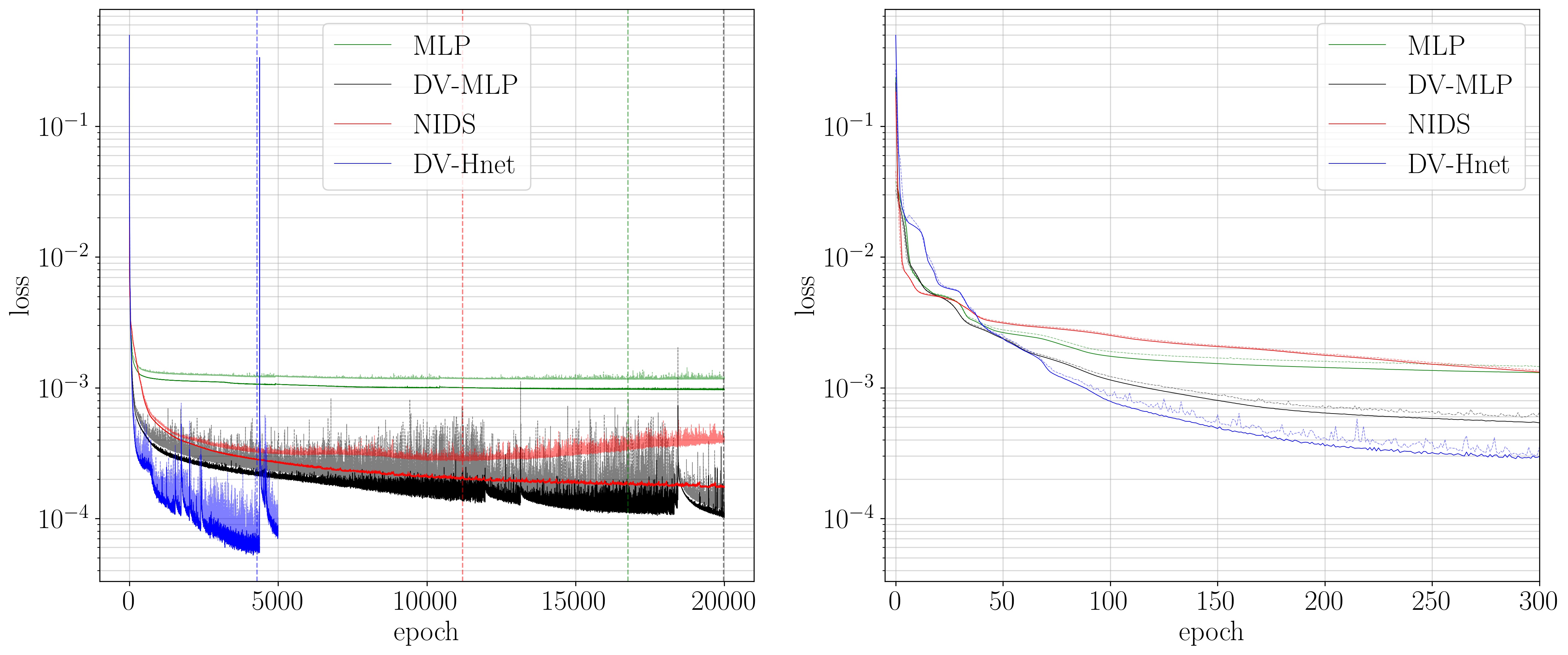}
	\caption{Comparison of training curves for the RANS problem, with detailed early behavior shown on the right.}
	\label{f:gm_training_curves_large}
\end{figure}

Models corresponding to the architectures given in Table \ref{t:model_architectures} were trained using 99 randomly selected cases in the training group and 25 cases in the validation group, corresponding to an 80/20 split.
As before, all models were trained with a learning rate of $1\times10^{-4}$, except for DV-Hnet which used a learning rate of $3\times10^{-5}$ to stabilize training behavior, and this is discussed further after other results are presented.
A batch size of 40,000 points was used, leading to 240 batches/optimizer updates per epoch.
A plot of the training curves is shown in Figure \ref{f:gm_training_curves_large}, where the transparent dashed curves correspond to validation loss, and vertical dashed lines locate the best validation loss, with weights from these locations used in the predictions which follow.
A summary of error metrics is given in Table \ref{t:gm_errors_1}.
As with the Poisson problem, training curves correspond to normalized outputs lying in the range $[0,1]$, while the errors reported in Table \ref{t:gm_errors_1} are dimensional.
Profiling results are given in Table \ref{t:gm_profile}, where mixed precision was utilized for all models.
The time-to-best-model is estimated using Equation \ref{eq:timetobestmodel}, and a full comparison of mixed and double precision training is given in Section \ref{s:profiling}. 
\FloatBarrier

\begin{table}[H]
	\begin{center}
		\caption{Summary of training and validation error metrics RMSE, MAE, and ML2E for the various models for the RANS problem.}
		\label{t:gm_errors_1}
		\begin{tabular}{|c|c|c|c|c|c|c|}
			\hline
			$\hat{\mb{q}}_{k}$  & Network Type  & RMSE (train / val)    & MAE (train / val)     & ML2E (train / val)\\
			\hline
			\multirow{4}{*}{$p$ [Pa]}            & MLP	        & 78.2 / 87.1    	    & 33.2 / 37.1        	& $2.97\cdot10^{-1}$ / $3.18\cdot10^{-1}$ \\
			                    & DV-MLP	    & 22.1 / 23.3           & 7.88 / 8.32        	& $8.32\cdot10^{-2}$ / $8.56\cdot10^{-2}$ \\
			                    & NIDS 	   	    & 27.9 / 37.0           & 9.57 / 18.8           & $1.05\cdot10^{-1}$ / $1.36\cdot10^{-1}$ \\
			                    & DV-Hnet 	    & 9.00 / 10.7           & 3.68 / 4.21        	& $3.42\cdot10^{-2}$ / $3.78\cdot10^{-2}$ \\
			\hline
			\multirow{4}{*}{$u$ [m/s]} 		    & MLP 	        & 3.00 / 3.22           & 1.64 / 1.80           & $1.24\cdot10^{-1}$ / $1.34\cdot10^{-1}$ \\
			        		    & DV-MLP 	    & 0.86 / 0.91           & 0.37 / 0.40           & $3.71\cdot10^{-2}$ / $3.91\cdot10^{-2}$ \\
			        		    & NIDS 	  	    & 1.13 / 1.37           & 0.51 / 0.71           & $4.86\cdot10^{-2}$ / $5.74\cdot10^{-2}$ \\
			         		    & DV-Hnet       & 0.62 / 0.66           & 0.22 / 0.24       	& $2.69\cdot10^{-2}$ / $2.82\cdot10^{-2}$ \\
			\hline
			\multirow{4}{*}{$v$ [m/s]} 		    & MLP	        & 1.86 / 2.03           & 0.57 / 0.62       	& $4.27\cdot10^{-1}$ / $4.55\cdot10^{-1}$ \\
			         		    & DV-MLP	    & 0.75 / 0.80           & 0.23 / 0.25       	& $1.72\cdot10^{-1}$ / $1.81\cdot10^{-1}$ \\
			         		    & NIDS 	   	    & 1.07 / 1.23           & 0.33 / 0.47       	& $2.46\cdot10^{-1}$ / $2.81\cdot10^{-1}$ \\
			         		    & DV-Hnet 	    & 0.40 / 0.49           & 0.12 / 0.14       	& $9.29\cdot10^{-2}$ / $1.02\cdot10^{-1}$ \\
			\hline
		\end{tabular}
	\end{center}	
\end{table}

\begin{table}[H]
	\begin{center}
		\caption{Comparison of maximum memory usage, average step times during training, and an estimated time-to-best-model for the RANS problem.}
		\label{t:gm_profile}
		\begin{tabular}{|c|c|c|c|c|}
			\hline
			Network Type    & Max. Memory & Average Step Time       & Epoch Best    & Time to Best Model \\
			\hline
			DV-MLP          & 1.06 GB      & 5.2 ms                  & 19,978       & 24,933 s   \\
			NIDS            & 1.13 GB      & 6.9 ms                  & 11,196       & 18,541 s   \\
			DV-Hnet         & 2.64 GB      & 31.0 ms                 & 4,293        & 31,940 s    \\ 
			\hline
		\end{tabular}
	\end{center}	
\end{table} 
As with the Poisson problem, the MLP model which does not consume the design variables performs considerably worse than the other models, though the gap is smaller here.
The training behavior is also quite different than what was observed for the Poisson problem, where NIDS and DV-Hnet models overfit the training data.
Here, a visible gap between the training and validation losses can be seen only for the NIDS model, while the curves are nearly coincident for DV-MLP and DV-Hnet models.

The separation and ranking can be seen clearly in the training curves and confirmed by Table \ref{t:gm_errors_1}.
The training-validation gap is larger for NIDS than for the others considered, in line with the training curves.
DV-Hnet clearly performs best, with reported MAEs roughly half the next-best model.
However, Table \ref{t:gm_profile} reveals that the time-to-best-model is greater for DV-Hnet than the others, despite convergence in a fewer number of epochs.
This difference is not extreme or proportional to the larger number of weights in the DV-Hnet model, or even to the difference in errors.
For example, the errors are roughly halved using DV-Hnet as compared to DV-MLP, but the time-to-best-model is only increased by roughly 30\%; an acceptable trade off.
DV-Hnet models do use more than twice the maximum memory of the other models.

\begin{figure}[h!]
	\centering
	\includegraphics[width=1.0\textwidth]{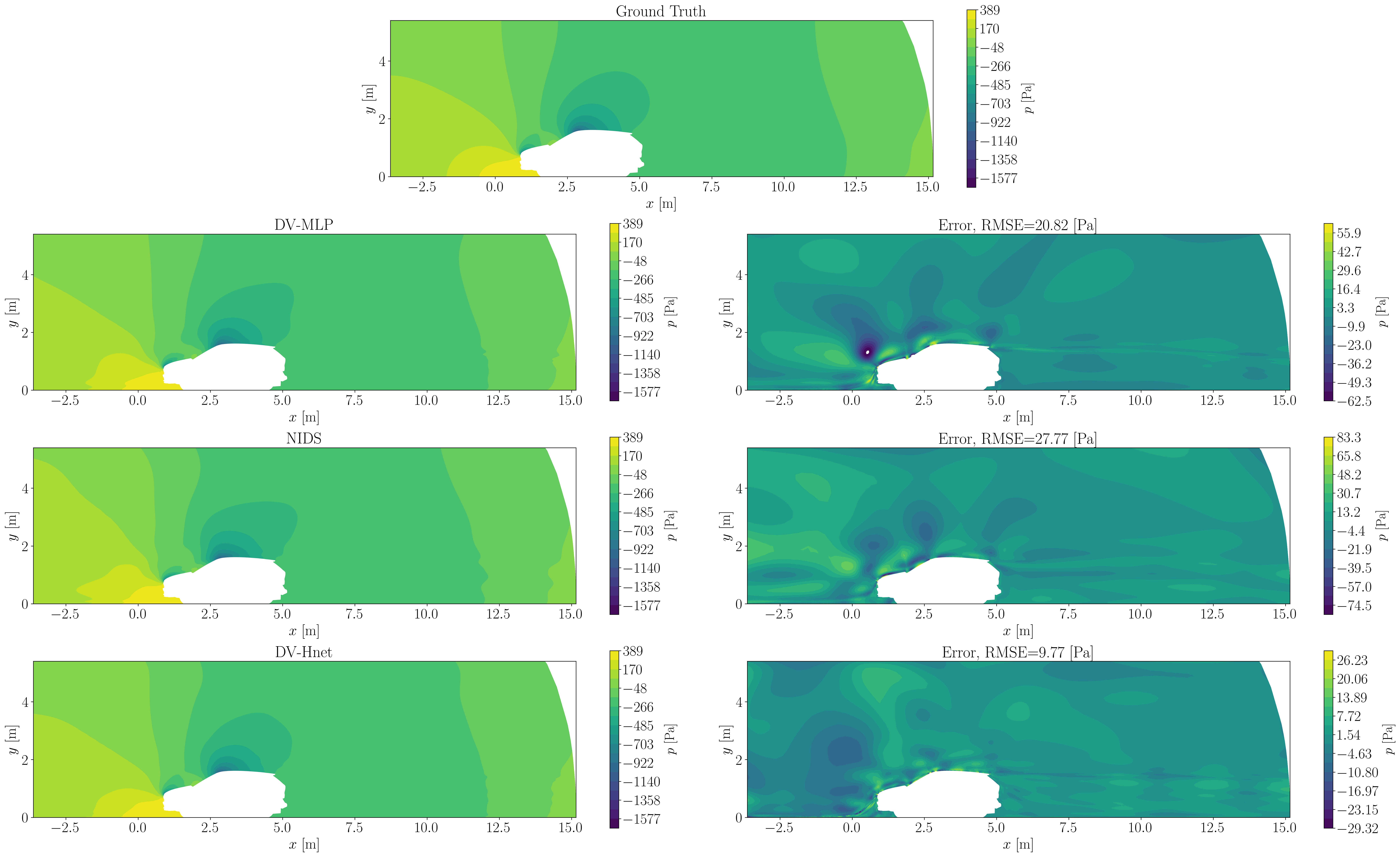}
	\caption{Comparison of predicted pressure fields, unseen vehicle shape.}
	\label{f:gm_pcontour}
\end{figure}

\begin{figure}[h!]
	\centering
	\includegraphics[width=1.0\textwidth]{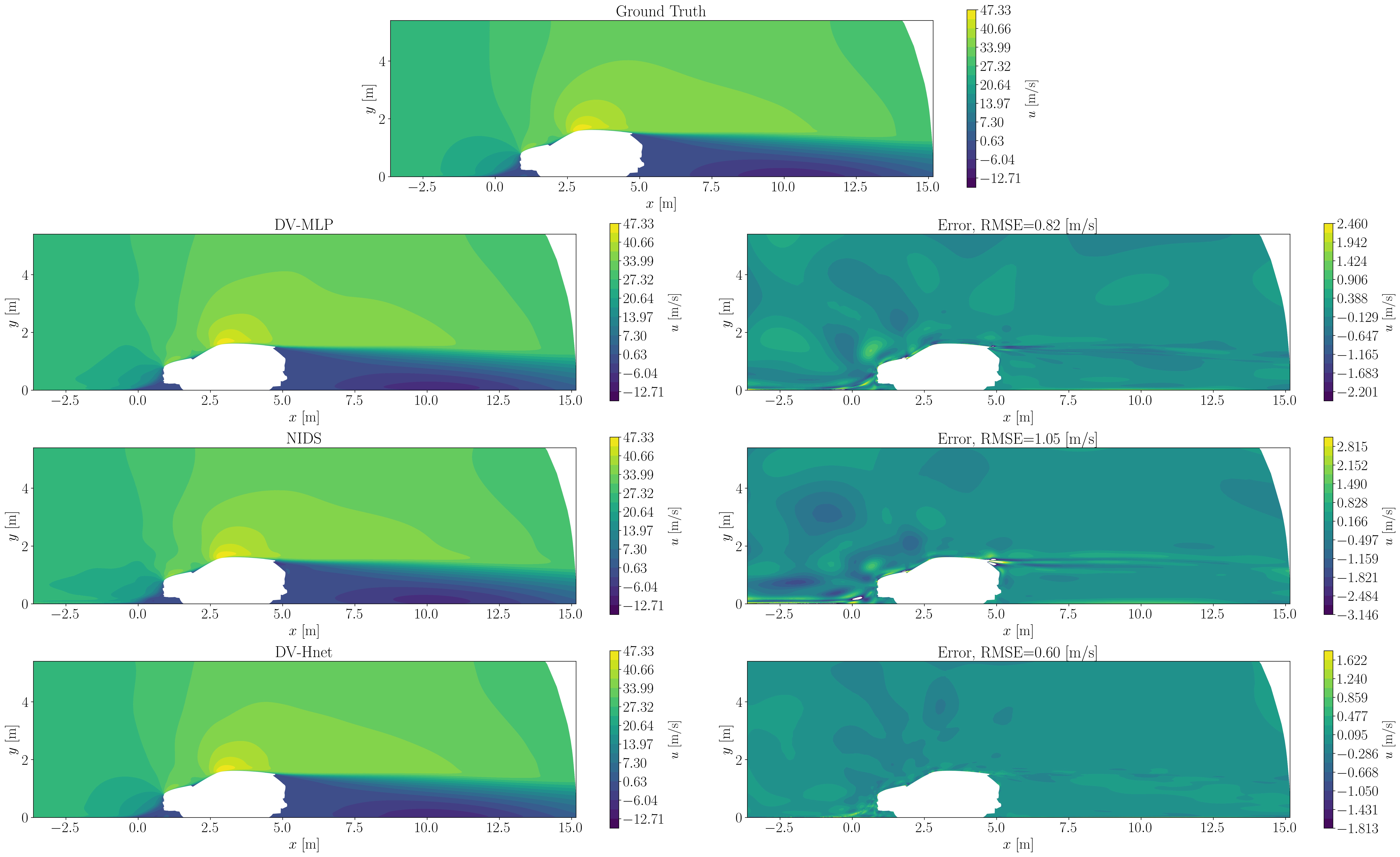}
	\caption{Comparison of predicted $x$-velocity fields, unseen vehicle shape.}
	\label{f:gm_ucontour}
\end{figure}

\begin{figure}[h!]
	\centering
	\includegraphics[width=1.0\textwidth]{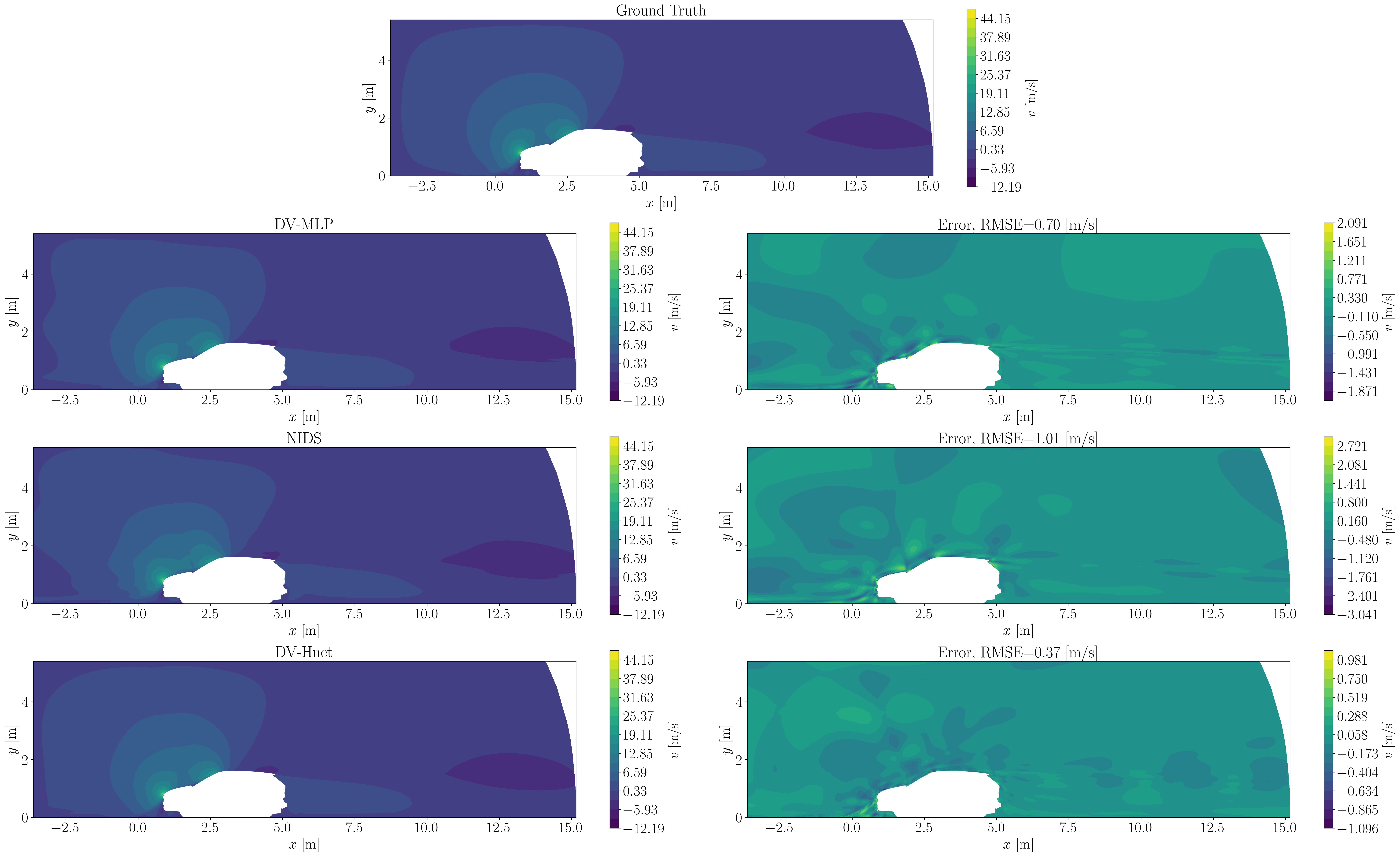}
	\caption{Comparison of predicted $y$-velocity fields, unseen vehicle shape.}
	\label{f:gm_vcontour}
\end{figure}

Comparisons of the ground truth and model predictions on an unseen vehicle shape are shown in Figures \ref{f:gm_pcontour}, \ref{f:gm_ucontour}, and \ref{f:gm_vcontour} for pressure, $x$-velocity, and $y$-velocity fields. 
The error colorbar scale is set to $\pm 3 \times \mathrm{RMSE}$ as small clusters of larger error overshadow finer details.
Regions with errors outside of this range are left white.
These figures show that all models capture the overall structure of the flow fields.
Figure \ref{f:gm_pcontour} shows that the high-pressure region in front of the vehicle and the low-pressure region on the vehicle roof are present in all predictions.
The lower-pressure vehicle wake and subsequent recovery are also captured.
Some distortion of the contour lines can be seen in the predictions, and are more prevalent for DV-MLP and NIDS than for DV-Hnet.
The DV-MLP error contours for pressure show two regions where the error is outside of the $\pm 3\times \mr{RMSE}$ range; in-front-of and above the hood, and along the windshield.
For all model predictions, areas of large error tend to exist close to the vehicle surface and within the fine details of the front grill.
\FloatBarrier
The $x$-velocity predictions of Figure \ref{f:gm_ucontour} reveal that the models capture the overall structure well, with small recirculating regions in front of the vehicle, acceleration and flow turning over the roof, and a decaying free-shear layer in the wake.
As before there are some discrepancies between the ground truth contour lines and the predictions, with the smallest deviations for DV-Hnet.
Small regions where the error is outside $\pm 3\times \mr{RMSE}$ can be seen for DV-MLP and NIDS, more prominently for NIDS.
As with the pressure, larger errors are frequently seen close to the vehicle surface and in the details of the front grill.

\begin{figure}[h!]
	\centering
	\includegraphics[width=1.0\textwidth]{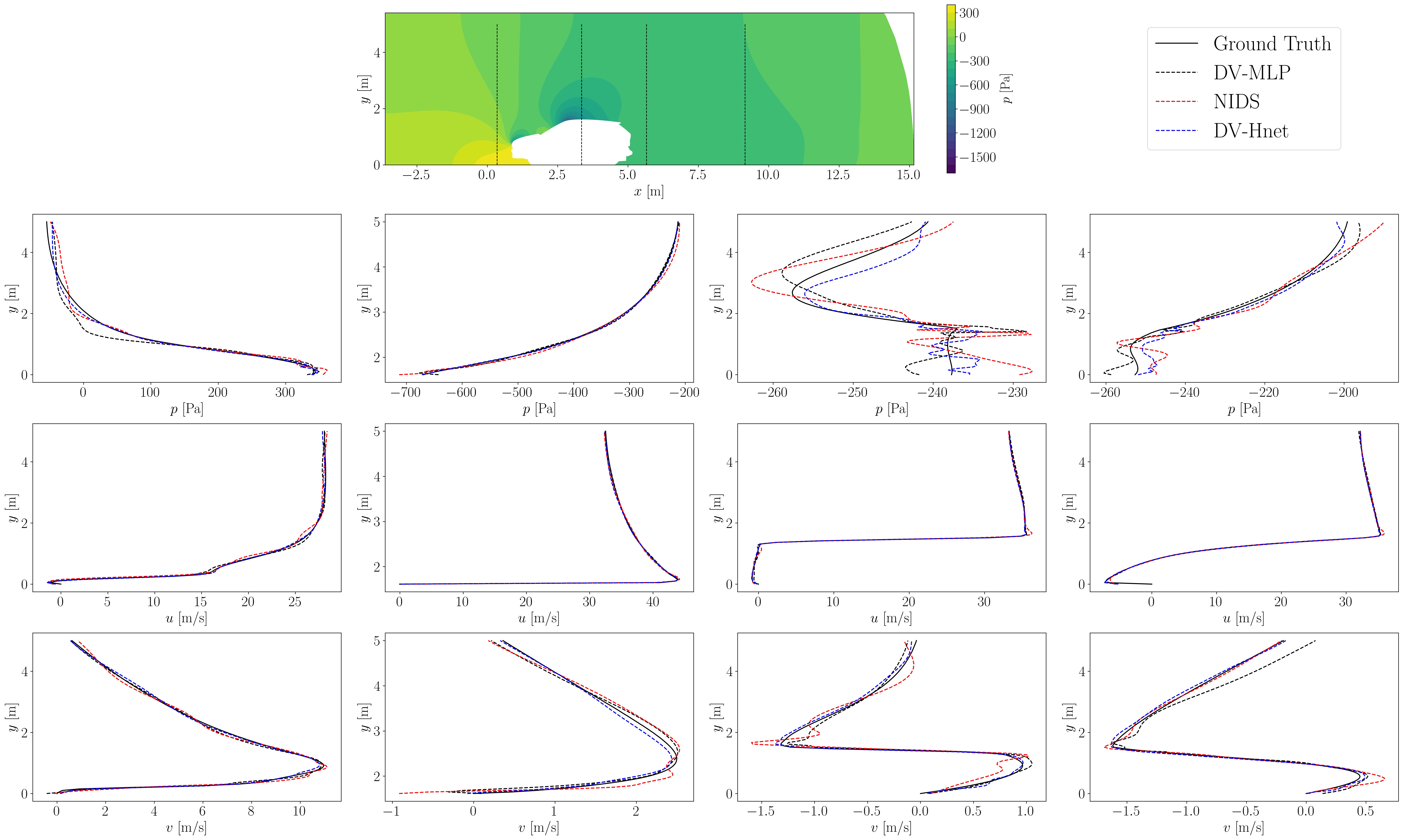}
	\caption{Comparison of line probes placed around the vehicle.}
	\label{f:gm_lineprobe}
\end{figure}

\FloatBarrier
Similar comments may be made regarding Figure \ref{f:gm_vcontour} for the $y$-velocity predictions, where all models capture the overall structure of the solution, but to differing degrees.
The dominant features of the $y$-velocity are the rapid vertical acceleration over the front of the hood and windshield, along with re-circulatory behavior in the wake.
Again, for all models the largest errors tend to cluster near the vehicle surface, especially in the front grill.
\FloatBarrier
For further comparison, vertical line probes are placed near the vehicle, with one in front, one through the vehicle's highest point, and two in the wake.
The probe in front of the vehicle is offset by 0.5 m, while those in the wake are offset by 0.5 m and 4 m.
The ground truth and model predictions at mesh points are interpolated to the line probe locations using the \texttt{griddata} function from the \texttt{scipy.interpolate} library, with the results shown in Figure \ref{f:gm_lineprobe}.
\FloatBarrier
Generally, the line probe predictions match the ground truth to a reasonable extent, though some oscillation is present in the network predictions.
This is most prevalent for the first pressure probe in the vehicle wake, though the effect is more pronounced due to the $x$-axis limits.

\begin{figure}[h!]
	\centering
	\includegraphics[width=1.0\textwidth]{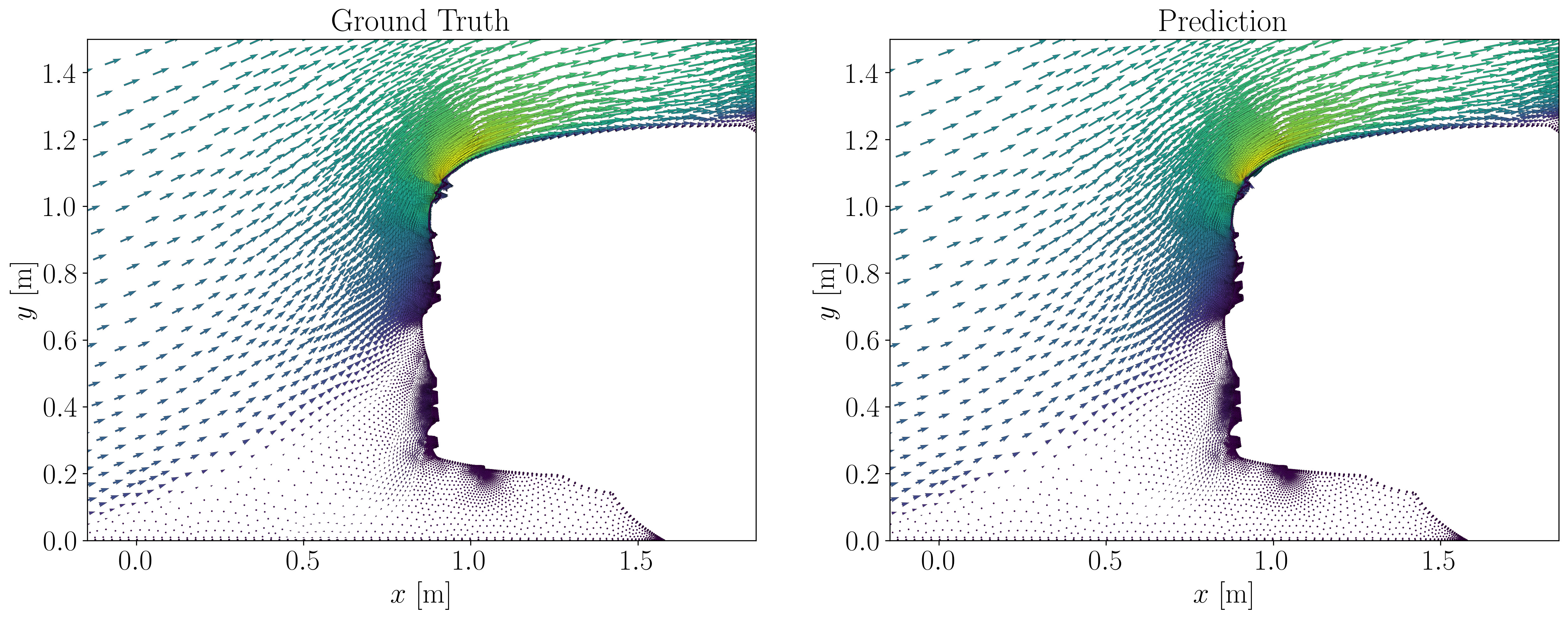}
	\caption{Comparing ground truth and DV-Hnet predicted velocity vectors near the front of an unseen vehicle shape. Differences are difficult to see by eye.}
	\label{f:gm_velvec_front}
\end{figure}

\begin{figure}[h!]
	\centering
	\includegraphics[width=1.0\textwidth]{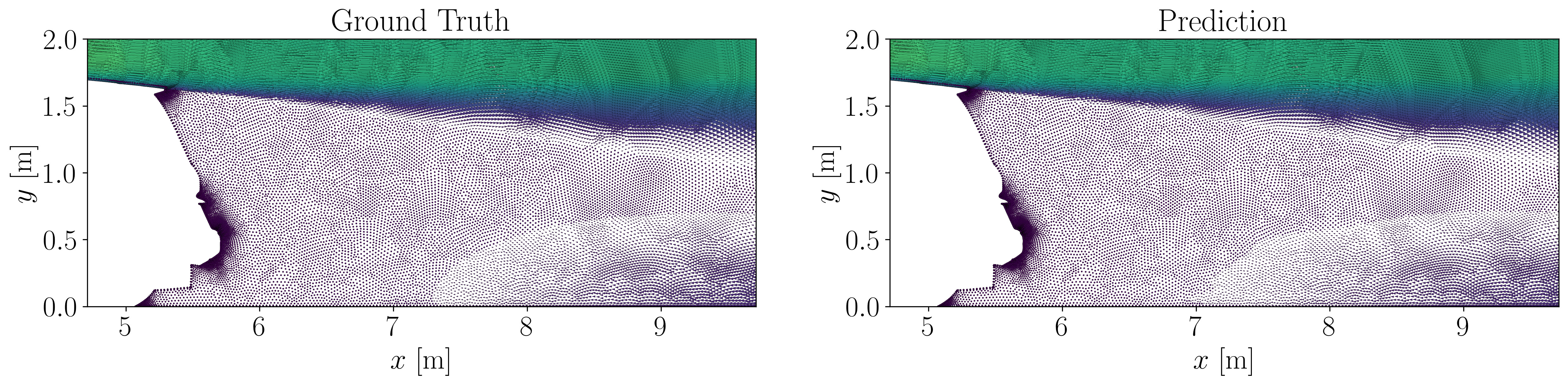}
	\caption{Comparing ground truth and DV-Hnet predicted velocity vectors in the wake of an unseen vehicle shape. Some discrepancies are seen in the recirculation zone. }
	\label{f:gm_velvec_rear}
\end{figure}

The ground truth and DV-Hnet predicted velocity vectors are shown in Figures \ref{f:gm_velvec_front} and \ref{f:gm_velvec_rear} for regions in front of and behind a different unseen vehicle shape.
In front of the vehicle, differences between the ground truth and predictions are difficult to see by eye.
The velocity vectors in the vehicle wake in Figure \ref{f:gm_velvec_rear} also match quite well, but some discrepancies in the shape and extent of the recirculation zone near the ground are visible.
\FloatBarrier
\FloatBarrier

In the discussion of the training curves, it was noted that DV-Hnet would become unstable during training when a learning rate of $1\times10^{-4}$ was used.
The unstable models would quickly converge, then jump back to and around higher loss values before the loss would become NaN or overflow to infinity.
This instability was observed more frequently when mixed-precision training was used, but was deemed an acceptable trade-off for the greater-than-10X reduction in training time, and it was found that reducing the learning rate helped to alleviate the issue, though instability is still seen in the training curve shown.
However, even unstable DV-Hnet models outperform DV-MLP and NIDS models, provided that the best model weights are tracked and retained during training.
These training issues may be due, in part, to the weight initialization schemes used, with all network weights initialized via the Glorot-uniform scheme \cite{glorot10a}.
While applications for hypernetworks abound, detailed study of their optimization, including principled weight initialization, is lacking.
This is noted in Ref. \cite{Chang2020Principled}, where the authors observed that use of the Glorot-uniform scheme in a hypernetwork produced weights which were improperly scaled for the main network, leading to instabilities in activations and losses.
They propose alternative methods for determining the width of the uniform distribution from which to draw initial weights for hypernetworks, termed Hyperfan-in and Hyperfan-out, and using these methods may help to stabilize the training dynamics. 

\subsection{Effect of Mixed-Precision Training}
\label{s:profiling}
Models for each problem were profiled using Tensorboard callbacks, with both mixed and double-precision training, with the average optimizer step-time and peak memory usage reported in Table \ref{t:profile}.
Mixed-precision training decreased the average step-time for all models for the RANS problem, but only DV-Hnet in the Poisson problem. 
The reasons for this are unclear, but likely related to the batch sizes of 1500 and 40,000 for the two problems, given that almost everything else is consistent between them.
The improvements due to mixed precision are most dramatic for DV-Hnet, with the average step time decreased by a factor of 10, and memory consumption cut by roughly a factor of 3 for the RANS problem.
These massive improvements allow DV-Hnet models to remain viable as compared to the other models, considering the time-to-best-model for each, despite the much greater model size.
For the RANS problem, the average step times for DV-MLP and NIDS were reduced by a factor of approximately 3 and 4.5 respectively, with only small reductions in memory consumption.
\begin{table}[H]
	\begin{center}
		\caption{Profiling metrics of average optimizer step time and peak memory usage compared for the various models, for mixed precision and double precision training.}
		\label{t:profile}
		\begin{tabular}{|c|c|c|c|c|c|c|c|c|}
			\hline
			 & \multicolumn{4}{|c|}{Average Step Time} & \multicolumn{4}{|c|}{Peak Memory Usage}\\
			 \hline
			 & \multicolumn{2}{|c|}{Poisson}    & \multicolumn{2}{|c|}{RANS}    & \multicolumn{2}{|c|}{Poisson} & \multicolumn{2}{|c|}{RANS} \\
			 \hline
			 Method     & Mixed     & Double   & Mixed     & Double      & Mixed        & Double        & Mixed     & Double     \\
			\hline
			DV-MLP      & 4.8 ms    & 2.9 ms    & 5.2 ms    & 14.7 ms    & 805 MB       & 812 MB      & 1.06 GB   & 1.25 GB     \\
			NIDS        & 6.5 ms    & 5.0 ms    & 6.9 ms    & 31.4 ms    & 807 MB       & 821 MB      & 1.13 GB   & 1.53 GB     \\
			DV-Hnet     & 5.8 ms    & 18.5 ms   & 31.0 ms   & 359.0 ms   & 874 MB       & 1.04 GB     & 2.64 GB   & 7.55 GB     \\
			\hline
		\end{tabular}
	\end{center}	
\end{table}

\section{Summary and Conclusions}
\label{s:conclusions}
The past few years have witnessed significant activity in the use of neural networks to develop surrogate representations of physical fields (e.g. ~\cite{Xu2020, Guo2016, Bhatnagar2019,Sekar2019}) on a given discretized mesh. 
The present work seeks the development of surrogate models on unseen meshes, mesh topologies, and geometries as a {\em continuous field}, allowing learning and prediction on meshes with arbitrary discretization and topology.
Three such models are proposed and compared; design-variable-coded multi-layer perceptron (DV-MLP), design-variable hypernetworks (DV-Hnet), and non-linear independent dual system (NIDS).
While each of these methods shares similarities to existing works, the details of the usage and implementations here are novel.
DV-MLP is a simple feed-forward neural network, with all input features fed through the main network.
DV-Hnet introduces an additional hypernetwork, generating all of the main-network weights and biases as a function of the design variables.
NIDS may be seen as a partial design-variable hypernetwork, with only the weights and biases of the linear output layer generated by the parameter network.
Input features include spatially-varying quantities, collected in $\mb{x}'$, and non-spatially-varying design variables, collected in $\pmb{\mu}$.
The spatial coordinates $\mb{x}$ are paired with a minimum-distance function evaluation $\phi(\mb{x};\pmb{\mu})$ to implicitly encode the geometry.

The proposed models were applied to a 2D Poisson problem defined with varying shapes embedded in a square domain with a spatially-distributed source term.
All models performed similarly on this problem as corroborated by the reported error metrics.
Good generalization performance was seen, with training and validation RMSE around 0.2 and MAE around 0.1, corresponding to mean-relative-errors around 0.1\% for all models.
However, the estimated time-to-best-model was much smaller for DV-Hnet than the others, despite the much greater number of model parameters.
This is due to the model converging in a fewer number of epochs, outweighing the larger average step-time.
It was observed that the models predicted significantly worse on triangles than on the other shape classes, and this is likely related to the incompleteness of design variable vector $\pmb{\mu}$, the lack of similarity between triangles and the other shapes, and the inability of the MDF to fully capture the problem geometry.

Models of each type were also trained on aerodynamic flows corresponding to 2D RANS solutions around complex vehicle shapes. 
The DV-Hnet model outperformed the others on this problem, with smaller RMSE and MAE by a factor of roughly 2, at a cost of about 30\% greater time-to-best-model, which we deem to be an acceptable trade.
However, training instability was observed for DV-Hnet, and without mixed-precision training it would not be competitive with DV-MLP and NIDS in terms of training time.
Interpreting the ML2E as a percentage error, DV-Hnet model predictions match the pressure field to within 4\%, the $x$-velocity to within 3\%, and $y$-velocity to within just above 10\%, with other models performing similarly, though worse.
The main features of the flow fields are well-captured by all models, with the largest errors often clustering in regions close to the vehicle and in the fine details of the grill.
Line probes showed some oscillation in the network predictions compared to the ground truth, consistent with discrepancies seen in the contour levels and velocity-vector plots showed that differences between prediction and ground truth are difficult to see by eye.

The results suggest that the proposed family of models can be accurate and effective alternatives to convolutional neural networks (CNNs) for surrogate modeling of PDE solution fields over complex geometries and arbitrary mesh topologies, with DV-Hnet models of primary interest.
It is again emphasized that CNNs typically require a fixed grid topology and have a large memory footprint for 3D problems since the entire grid is an input.
In contrast, the present approaches take pointwise information and design variables as inputs, allowing the size of the model and the memory requirements to be decoupled from the solution field degrees of freedom. 
Further study of the hypernetwork training process and weight initialization scheme are warranted, and explorations of alternate hypernetwork architectures which reduce the model parameter count are of particular interest.

\section*{Acknowledgments}
{This work is funded by General Motors, Inc. under a contract titled “Deep Learning and Reduced Order Modeling for Automotive Aerodynamics,”  and by Advanced Research Projects Agency-Energy (ARPA-E) DIFFERENTIATE program under the project ``Multi-source Learning-accelerated Design of High-efficiency Multi-stage Compressor,'' in collaboration with Raytheon Technologies Research Center (RTRC). Computing resources were provided by the NSF via grant 1531752 MRI: Acquisition of Conflux, A Novel Platform for Data-Driven Computational Physics. Some contents of this paper have appeared online in an unpublished preprint \cite{duvall2021nonlinear}. }

\bibliographystyle{elsarticle-num}
\bibliography{refs.bib}
\section{Appendix}
\label{s:appendix}

\subsection{Neural Network Implementation Details}
\label{s:training_methods}
\paragraph{Model Implementation and Training}
All models are implemented via Python 3.X classes using Tensorflow v2.X \cite{abadi2016tensorflow} and are trained using a Nvidia RTX A6000 48 GPU.
The DV-MLP implenentation uses off-the-shelf Tensorflow-Keras sequential models, constructed using the Keras functional API.
The DV-Hnet and NIDS models subclass Tensorflow-Keras Models (\texttt{tf.keras.Model}), overwriting the \texttt{call} method appropriately.
The DV-Hnet implementation uses a Tensorflow-Keras sequential model for the hypernetwork, required during class instantiation.
Similarly, NIDS models use Tensorflow-Keras sequential models for both the spatial and parameter networks.
Tensorflow-Keras models are useful as they provide high-level abstraction and contain easy-to-use functions for common tasks, such as training models and saving/loading network weights.
All model weights are initialized using the Glorot-uniform weight-initialization scheme \cite{glorot10a}, and trained using calls to \texttt{tf.keras.Model.fit()}.
Adam optimizer with default settings is used with a learning rate of $1\times10^{-4}$ for all numerical experiments, unless otherwise noted.
Mixed precision is used in training all models, and model checkpoints are used to save the model weights corresponding to the best training and validation losses obtained as training progresses.

A simple mean-squared-error loss function is used in all cases, without any kernel or activity regularization.
For DV-Hnet and NIDS models, two different approaches to handling batching are possible.
\begin{itemize}
	\item \textbf{Method 1: Batch by Case} Mini-batches are created consisting of only points for a single case $j$ with the same parameter vector $\pmb{\mu}^{j}$. 
	The output of a single forward-pass of the hypernetwork is combined with multiple forward-passes of the main/spatial network(s) for all points in the mini-batch, followed by an optimizer update. 
    This is more computationally efficient than evaluating the parameter network for each spatial point.
    \item \textbf{Method 2: Mixed Batches} Mini-batches are created which consist of points from different cases, and the hypernetwork and main/spatial networks are forward-propagated for each data point. 
    The parameter network input vector $\pmb{\mu}^{j}$ is tiled $n_{j}$ times for each case. 
    With this method there is greater computational overhead due to the much larger number of hypernetwork calls per mini-batch. 
    \end{itemize}
Regardless of the batching used, the loss may be expressed as
\begin{equation}
	\label{eq:loss}
	\mc{L}(\theta) = \frac{1}{N}
	\sum_{j=1}^{n_{c}} \sum_{m=1}^{n_{j}} \| \hat{\tilde{\mb{q}}}(\mb{x}_{m}^{j}, \pmb{\mu}^{j};\theta) - \tilde{\mb{q}}(\mb{x}_{m}^{j}, \pmb{\mu}^{j}) \|_{2}^{2},
\end{equation}
where $N$ is the total number of mesh points in all cases, written as
\begin{equation}
	N = \sum_{j=1}^{n_{c}} n_{j}.
\end{equation}
There are advantages and disadvantages to each of these methods. 
For method 1, the stochasticity is decreased as compared to fully-mixing the data for all cases as with method 2. 
Increased stochasticity in training is qualitatively seen to help with convergence and freeing the network from local minima in the high-dimensional network-weight-space.
Additionally, since each solution may have a different number of points a separate computational graph must be built for each case when method 1 is used, causing the first epoch to take a comparatively long time to complete and raising the memory and computational requirements to build and store all graphs. 
Subsequent epochs run more quickly once all graphs are built. 
When method 2 is used the minibatch size may be fixed so fewer computational graphs are required. 
A trade-off exists between the two relating to convergence, computational cost and overhead, though the details of this are not fully explored here. 
Method 2, mixed batches is used in training all networks reported in the main text, and the models in the repository are hard coded to use this method to call the model in training.

\paragraph{Normalization}
All inputs and outputs are min-max normalized using the statistics of the training group, on a signal-by-signal basis, so that they lie approximately in the range $[0,1]$. Some members of the validation group may be slightly above or below this range if they are smaller than the smallest element of the training set or larger than the largest element of the training set. Vectors $\mb{x}/\pmb{\mu}/\mb{q}$ are normalized component-wise. The formula for computing the normalization is 
\begin{equation}
	\label{eq:range_norm}
	\tilde{r}^{j} = \frac{r^{j} - \min(\mb{r})}{\max(\mb{r}) - \min(\mb{r})},
\end{equation}
where $r$ is an element of $\mb{x}$, $\mb{q}$, or $\pmb{\mu}$ from either the training or validation group. Vector $\mb{r}$ is the collection of all instances of $r$ from the training dataset, $r^{j}$ is dimensional, and $\tilde{r}^{j}$ is the normalized quantity. The predictions are re-dimensionalized for computing errors and plotting by rearranging Equation \ref{eq:range_norm} for $r^{j}$.

\subsection{2D Poisson Dataset, Solution Details}
\label{s:appendix_poisson}
Triangular meshes are generated for each random shape using the Gmsh Python API \cite{Geuzaine2009}, though the surface mesh of each shape are fully specified programatically such that the distance between adjacent surface nodes, $\Delta r$, is approximately $5\times 10^{-3}$.
For polygons, the coordinates of the vertices are given as initial mesh points. 
Then the line connecting adjacent vertices is divided into $n$ equal segments, where $n$ is selected as the smallest integer such that the distance between mesh points is less than or equal to $5\times 10^{-3}$. 
For circles, the number of points is chosen by rearranging the (approximate) arc length formula $\Delta s = r \Delta\gamma$ to compute the required $\Delta \gamma$ while setting $\Delta s =5\times 10^{-3}$. Then the number of points is chosen as $n =  \bigg( \frac{2 \pi}{\Delta \gamma}     \bigg)$.
The mesh for the rest of each domain $\Omega^{j}$ is generated using the Gmsh python API with mesh spacing on $\partial B$ set to $5\times 10^{-2}$. 
The meshes are saved to .vtk format and the governing equations solved using the finite element method using SfePY \cite{Cimrman_Lukes_Rohan_2019}.

\subsection{NIDS: Further Details}
\subsubsection{Model Mappings}
\label{s:appendix_nids}
Functionally, the mapping of the spatial network is
\begin{eqnarray}
	\label{eq:spatial_map}
	N_{x}&:& \mc{X}' \rightarrow \mc{H},
\end{eqnarray}
the mapping of the parameter network is
\begin{eqnarray}
	\label{eq:param_map}
	N_{\mu}&:& \mc{M} \rightarrow \mc{W},
\end{eqnarray}
and that of the linear output layer may be written as
\begin{eqnarray}
	\label{eq:output_map_1d}
	N_{o}&:& \mc{W}\times \mc{H} \rightarrow \mc{Q}.
\end{eqnarray}
We then note that the triple $(\mc{W}, \mc{H}, N_{o})$ defines a dual system over a subset of the real numbers $\mathbb{R}$ when the system state $\mb{q}$ is one-dimensional.
When the system state is $n$-dimensional, then $n$ dual systems are induced of the form $(\mc{W}_{i}^{'}, \mc{H}, N_{o,i})$, with bilinear map $N_{o,i}$ described as
\begin{eqnarray}
	\label{eq:output_map_nd}
	N_{o,i}&:&\mc{W}_{i}^{'} \times \mc{H} \rightarrow \mc{Q}_{i}^{'},
\end{eqnarray}
which corresponds to the inner product of each row of $\mb{W}_{\mu}$ with $\mb{h}_{x}$, and addition of the bias term. 
In this case, $\mc{W}_{i}^{'} \subset \mathbb{R}^{n_{h}+1}$ and $\mc{Q}_{i}^{'} \subset \mathbb{R}$, corresponding to the $i$th dimension of the system state. 
We note that the maps are bilinear in $\mb{w}_{\mu}$ and $\mb{h}_{x}$, not in $\mb{x}$ and $\pmb{\mu}$ due to the non-linear nature of neural networks. 

From this we can see that the name `non-linear independent dual system' is an apt description of how the model output layer operates. 
That is, the tripe describing the output layer $(\mc{W}, \mc{H}, N_{o})$ is a dual system or induces multiple dual systems. 
The vectors from each space $\mb{w}_{\mu} \in \mc{W}$ and $\mb{h}_{x} \in \mc{H}$ serve as inputs to the dual system bilinear map $N_{o}$, and are generated non-linearly and independently by the neural networks.

\subsubsection{Modal Interpretation of NIDS Predictions}
\label{s:NIDS_modes}
Proper orthogonal decomposition (POD) and dynamic mode decomposition (DMD) are popular techniques originally developed for analysis of flow fields and turbulence \cite{Lumley1967pod,Schmid2010}, and are now frequently used for model order reduction, prediction, and control \cite{benner2015survey,proctor2016dynamic,salmoiraghi2018free}. Here we examine POD reconstruction, then show how NIDS predictions may be interpreted in a similar manner.  

Consider $n$ solution snapshots of a system with a single state variable defined on a $d$-dimensional common mesh with $m$ nodes, where each solution corresponds to a unique set of design variables $\pmb{\mu}$. 
The solution at a spatial location $i$ is written as $q(\mb{x}_{i},\pmb{\mu}^{j})$, and a snapshot of the state at all locations may be written as $\mb{q}^{j} \in \mathbb{R}^{m}$ or compactly as $\mb{q}$. 
Note that elsewhere in this work $\mb{q}$ represents a multi-dimensional state, where in this section it is a snapshot-vector of the state at all mesh locations. 
Additionally, collect mesh spatial-coordinates in matrix $\mb{X} \in \mathbb{R}^{m \times d}$ and all solution snapshots in data matrix $\mb{Q} \in \mb{R}^{m\times n}$. To get the POD basis, one may perform a singular-value-decomposition on data matrix $\mb{Q}$, written as
\begin{eqnarray}
	\label{eq:pod_svd}
	\mb{Q} &=& \mb{U\Sigma V}^{*}.
\end{eqnarray}
Then select the first $r$ singular vectors to form basis matrix
\begin{eqnarray}
	\label{eq:pod_basis}
	\mb{\Phi} &=&  \mb{U}[:,1:r] \in \mathbb{R}^{m \times r},
\end{eqnarray}
which has orthonormal columns commonly called modes. Let $\pmb{\phi}_{i}$ represent the $i$th POD mode. 

A snapshot from the collection may be reconstructed using basis $\mb{\Phi}$ by first computing the basis coefficients $\mb{c}$ as
\begin{equation}
	\label{eq:pod_coeff}
	\mb{q} \approx \mb{\Phi} \mb{c} \rightarrow \mb{c} = \mb{\Phi}^{T}\mb{q}.
\end{equation} 
This takes advantage of the orthonormality ($\mb{\Phi}^{T}\mb{\Phi} = \mb{I}_{r}$) of POD modes. 
The resulting POD projection of the snapshot is then $\hat{\mb{q}} = \mb{\Phi c}$.  
Additionally, we note that each POD mode is a function of the mesh coordinates in $\mb{X}$ and the basis coefficients are dependent on the parameters $\pmb{\mu}$ associated with snapshot $\mb{q}$. 
Then the POD projection may be written as
\begin{equation}
	\label{eq:pod_summation}
	\hat{\mb{q}}(\mb{X}, \pmb{\mu})_{\mr{POD}} = \mb{\Phi}\mb{c} = \sum_{i=1}^{r} c_{i}(\pmb{\mu})\pmb{\phi}_{i}(\mb{X}).
\end{equation}
Next consider the approximation of a snapshot $\mb{q}$ with a NIDS network, where a prediction for location $i$ is 
\begin{equation}
	\label{eq:pod_nids_pred}
	\hat{q}(\mb{x}_{i}, \pmb{\mu}) = \mb{W}_{\mu}(\pmb{\mu}) \mb{h}_{x}(\mb{x}_{i}) + b_{\mu},
\end{equation} 
where $\mb{x}_{i}$ is the $i$th row of $\mb{X}$.
The full snapshot is attained by stacking the NIDS predictions at all locations. The system state is 1D, so $\mb{W}_{\mu} \in \mathbb{R}^{1 \times n_{h}}$. 
Re-express this as a vector instead of a matrix as $\mb{w}(\pmb{\mu}) = \mb{W}_{\mu}^{T}$, noting that $\mb{w}(\pmb{\mu}) \neq \mb{w}_{\mu}$ as is defined as the output of the parameter network in Equation \ref{eq:param}. 
The stacked NIDS approximation is then
\begin{equation}
	\hat{\mb{q}}(\mb{X}, \pmb{\mu})_{\mr{NIDS}} = \begin{bmatrix} \mb{w}(\pmb{\mu})^{T} \mb{h}_{x}(\mb{x}_{1}) + b_{\mu}(\pmb{\mu})  \\
		\mb{w}(\pmb{\mu})^{T} \mb{h}_{x}(\mb{x}_{2}) + b_{\mu}(\pmb{\mu})  \\
		\vdots \\
		\mb{w}(\pmb{\mu})^{T} \mb{h}_{x}(\mb{x}_{m}) + b_{\mu}(\pmb{\mu})  \\
	\end{bmatrix}.
\end{equation}
Rewrite the NIDS approximation as
\begin{eqnarray}
	\label{eq:NIDS_decomp}
	\hat{\mb{q}}(\mb{X}, \pmb{\mu})_{\mr{NIDS}} &=&  
	\begin{Bmatrix}
		\mb{w}(\pmb{\mu})^{T}
		\begin{bmatrix}
			h_{x,1}(\mb{x}_{1}) & h_{x,2}(\mb{x}_{1}) & \cdots & h_{x,n_{h}}(\mb{x}_{1}) \\
			h_{x,1}(\mb{x}_{2}) & h_{x,2}(\mb{x}_{2}) & \cdots & h_{x,n_{h}}(\mb{x}_{2}) \\ 
			\vdots  & \vdots & \ddots & \vdots \\
			h_{x,1}(\mb{x}_{m}) & h_{x,2}(\mb{x}_{m}) & \cdots & h_{x,n_{h}}(\mb{x}_{m}) \\ 
		\end{bmatrix}^{T}
	\end{Bmatrix}^{T}
	+ \begin{bmatrix}
		b_{\mu}(\pmb{\mu}) \\
		b_{\mu}(\pmb{\mu}) \\
		\vdots \\
		b_{\mu}(\pmb{\mu}) 
	\end{bmatrix}\\
	\label{eq:NIDS_decomp2}
	&=& \big( \mb{w}^{T} \mb{H}^{T} \big)^{T} + \mb{b}_{rep} \\
	\label{eq:NIDS_decomp3}
	&=& \mb{H}\mb{w} + \mb{b}_{rep}\\
	\label{eq:nids_modes}
	\hat{\mb{q}}(\mb{X}, \pmb{\mu})_{\mr{NIDS}} &=& \sum_{i=1}^{n_{h}} w_{i}(\pmb{\mu})\mb{h}_{i}(\mb{X}) + \mb{b}_{rep}(\pmb{\mu}),
\end{eqnarray}
where these expression define $\mb{H} \in \mathbb{R}^{m \times n_{h}}$ and $\mb{b}_{rep} \in \mathbb{R}^{m \times 1}$ by comparison between Equations  \ref{eq:NIDS_decomp} and \ref{eq:NIDS_decomp2}. 
Simplifying and re-expressing the prediction in this way leads to Equation \ref{eq:nids_modes}, where $\mb{h}_{i}$ is the $i$th column of $\mb{H}$, corresponding to the $i$th entry of $\mb{h}_{x}$ collected for all spatial locations in $\mb{X}$. 

Equation \ref{eq:nids_modes} for the NIDS prediction has a very similar structure as Equation \ref{eq:pod_summation} for the POD reconstruction, except for the additional bias term and the fact that the NIDS network may also consume an additional MDF/SDF coordinate in $\mb{x}$. 
However $\mb{X}$ may be defined to include this additional coordinate without affecting the POD analysis. 
Thus $\mb{h}_{i}$ may also be interpreted as a spatial mode and plotted/analyzed in a similar way as a POD mode. 
A key difference is that the POD modes were obtained via decomposition of a snapshot matrix, requiring a consistent discretization, while NIDS operates pointwise and the modes are learned during training. 

Equation \ref{eq:nids_modes} inverts the role of the spatial and parameter networks as previously described.
Previously the parameter network was conceptualized as providing a basis which applies globally for a solution instance, while the spatial network provides coordinates relative to that basis to generate the prediction at a spatial location.
Whereas with the above interpretation, the coordinates provided by the spatial network are collected for all locations and interpreted as basis functions, with the parameter network now providing the coordinates for the global modes.
The POD modes have additional structure beyond orthonormality since they are obtained via SVD, and thus the modes have an inherent order defined by the singular values. 
And more critically, it may be shown that that the POD reconstruction of data matrix $\mb{Q}$ is the optimal rank $r$ reconstruction in a Frobenius norm sense. 
NIDS modes lack these structures and this guarantee, but indeed there are possibilities to impose such structure by including appropriate penalizing terms in the loss function during training.
\end{document}